\documentclass[aps,prl,twocolumn,a4paper,10pt,notitlepage,footinbib,superscriptaddress,longbibliography]{revtex4-1}
\usepackage[english]{babel}
\usepackage{lmodern}
\usepackage[latin9]{inputenc}
\usepackage{endnotes}
\usepackage{amssymb,amsmath,amsfonts}
\usepackage{textcomp}
\usepackage{braket}
\usepackage{soul}

\usepackage{graphicx}% Include figure files
\usepackage{dcolumn}% Align table columns on decimal point
\usepackage{bm}% bold math
\usepackage{xcolor}
\usepackage{soul}

\begin{document}

\title{Spontaneous motion of a passive fluid droplet in an active microchannel}

\author{A. Tiribocchi$^*$} 
\affiliation{Istituto per le Applicazioni del Calcolo CNR, via dei Taurini 19, 00185 Rome, Italy\\ Corresponding author: adrianotiribocchi@gmail.com}
\author{M. Durve}
\affiliation{Center for Life Nano Science@La Sapienza, Istituto Italiano di Tecnologia, 00161 Roma, Italy}
\author{M. Lauricella}
\affiliation{Istituto per le Applicazioni del Calcolo CNR, via dei Taurini 19, 00185 Rome, Italy}
\author{A. Montessori}
\affiliation{Department of Engineering, Universit\`{a} degli Studi Roma Tre, Via Vito Volterra 62, 00146 Rome, Italy}
\author{S. Succi}
\affiliation{Istituto per le Applicazioni del Calcolo CNR, via dei Taurini 19, 00185 Rome, Italy}
\affiliation{Center for Life Nano Science@La Sapienza, Istituto Italiano di Tecnologia, 00161 Roma, Italy}
\affiliation{Department of Physics, Harvard University, Cambridge, MA, 02138, USA}

\begin{abstract}
We numerically study the dynamics of a passive fluid droplet confined within a microchannel whose walls are covered with a thin layer of active gel. The latter represents a fluid of extensile material modelling, for example, a suspension of cytoskeletal filaments and molecular motors. Our results show that the layer is capable of producing a spontaneous flow triggering a rectilinear motion of the passive droplet. For a hybrid design (a single wall covered by the active layer), at the steady state the droplet attains an elliptical shape, resulting from  an asymmetric saw-toothed structure of the velocity field. On the contrary, if the active gel covers both walls, the velocity field exhibits a fully symmetric pattern considerably mitigating morphological deformations. We further show that the structure of the spontaneous flow in the microchannel can be controlled by the anchoring conditions of the active gel at the wall. These findings are also confirmed by selected 3D simulations. Our results may stimulate further research addressed to design novel microfludic devices whose functioning relies on the collective properties of active gels.
\end{abstract}

\maketitle

\section{Introduction}

The last years have seen a rapid surge in the study of active matter, which deals with systems composed of self-driven units capable of converting stored energy into systematic movement \cite{marchetti}. Active matter encompasses a large variety of natural and artificial instances, ranging from flocks of birds \cite{parisi} and school of fishes \cite{okubo} at the macroscopic scale to suspensions of bacteria \cite{dombrowski}, cytoskeletal proteins \cite{joanny} and self-propelled colloids at the microscopic one \cite{bechinger}.  Their inherent non-equilibrium nature results in a wealth of enthralling phenomena, such as spontaneous flows \cite{surrey,ramin2}, active turbulence \cite{wensink,dunkel}, anomalous diffusion \cite{ramin}, superfluid-like behavior \cite{clement,saintillan} and motility induced phase separation \cite{mips,gonnella}, to name a few ones.

Innovative examples of autonomous systems are synthetic self-propelled droplets, i.e.  spatially confined fluid suspensions whose motion is guided by sophisticated modes of propulsion mechanisms which can involve, for example, the use of an active gel confined within \cite{dogic,sagues2,sagues3,rao,maeda}.  Some of the best known experimental realizations of such active materials are actomyosin solutions \cite{koler,loisel} and suspensions of microtubules and kinesin \cite{surrey,sumino}, which are soft fluids comprising force dipoles exhibiting a long range orientational order typical of liquid crystals \cite{sriram,ramaswamy,shendruk}. 
The former is an example of contractile material since the dipolar forces (exerted by the myosin) are directed towards the center of mass, whereas the latter pertains to extensile fluids. From a technological perspective, active gel droplets are gaining significant interest as model systems for studying the dynamics of micro-organisms, such as cells \cite{tjhung1,giomi,gareth_pre,zwicker,yeom_nat}, and for designing artificial microswimmers \cite{tiribocchi_nat,carenza_pnas,liverpool_prl,maass,maass2,aranson1,aranson2,aranson3,ruske}.

Alongside these objects, an alternative design is represented by the "inverted" counterpart,  i.e. passive Newtonian droplets surrounded (partially or entirely) by an active (or passive) gel \cite{demagistris,poulin}. A number of theoretical and experimental works has demonstrated that the inclusion of micron-sized objects in an active fluid bath often results in a persistent motion induced by a combination of mechanical energy extracted by the surrounding fluid and a careful manufacturing of the object, such as rigid structures with asymmetric boundaries \cite{dileonardo,dileonardo2,sokolov} or soft deformable ones, like polymeric chains \cite{nikola,joakim}.  It would be then natural to ask whether similar conditions can be realized in the context of inverted liquid crystals, where a passive drop can be spontaneously set into motion using the energy provided by the external active gel. Indeed, while many efforts have been dedicated in realizing self-motile droplets containing the active material, a much less investigated system is that in which the latter is hosted in the exterior environment.

In this paper we consider a microchannel comprising a mixture of an active extensile gel and a passive isotropic fluid, where the former is confined within a thin layer attached to the wall while the latter occupies the rest of the channel and is made of a Newtonian fluid droplet immersed in a second fluid (see Fig.\ref{fig1}a). With respect to a typical "inverted" design  where a passive drop would be fully embedded in an active medium, here the active component is distanced from the drop and occupies only a small portion of the channel. Such a system could be  realized in the lab by dispersing self-attractive cytoskeletal gels  in a layer of water wetting the wall (following a protocol akin to that discussed in Ref.\cite{dogic}), while the passive mixture could be produced through usual microfluidic techniques \cite{maass,petit,diota}.

We will show below that this geometry is capable of producing a spontaneous flow triggering a unidirectional motion of the passive drop, with a speed strongly depending on the activity. We also find that, if the active layer covers a single wall, the flow propelling the drop displays an asymmetric saw-tooth profile, fostering significant morphological changes for high activities. Alternatively, if the layers cover both wall, the fluid velocity acquires a symmetric profile suppressing droplet deformations. In agreement with previous works \cite{mare_yeom}, our results also suggest that controlling the orientation of the active gel at the boundaries decisively affects the emerging  flow as well as speed and shape of the droplet. These findings could be relevant for the design of autonomous microdevices able to sort and drive non-motile drops in a predictable manner, a challenging task in active matter especially if the need of keeping a certain direction of motion is required for long periods of time\cite{marchetti3,lavrentovich}. In addition, they may be potentially useful for realizing energy-saving soft systems whose functioning would require a reduced amount of active material (and hence energy) with respect to a fully inverted geometry \cite{demagistris}.

The paper is structured as follows. In the next section we shortly describe the theory and some numerical details while in the following ones we illustrate the results. First we discuss the physics of passive drop plus a single active layer in terms of structure of the flow and droplet kinematics, and then we move on to the case with two layers. Afterwards, the effect of boundary conditions is described and, before concluding, a selection of 3D results is presented.

\section{The Model}

Here we shortly summarize the theoretical model used in this work. Our system consists of a two dimensional microfluidic channel in which a passive fluid droplet is surrounded by a passive solvent (see Fig.\ref{fig1}a) while a layer of extensile material is glued at the bottom wall. Its physics is described using a combination of phase field modeling and active liquid crystal hydrodynamics, an approach already adopted in previous works \cite{tiribocchi2,tiribocchi_nat,tir_pre,pof1,pof2}. More specifically, we consider two scalar phase fields $\phi_i({\bf r},t)$ ($i=1,2$) accounting for the density of the droplet and the layer, a polar field ${\bf P}({\bf r},t)$ capturing the mesoscale orientation of the active material confined within the layer, the density $\rho({\bf r},t)$ of the fluid and its global velocity ${\bf v}({\bf r},t)$.  The presence of a further active component (such an extensile layer on the top wall) would require the inclusion of additional scalar and polar fields.    

The equilibrium features of a passive suspension are encoded in the following free energy density
\begin{eqnarray}\label{free}
f&=&\frac{a}{4\phi_{cr}^4}\left[\phi_1^2(\phi_1-\phi_0)^2+\phi_2^2(\phi_2-\phi_0)^2\right]\nonumber\\
&&+\frac{k}{2}\left[(\nabla\phi_1)^2+(\nabla\phi_2)^2\right]+\epsilon\phi_1\phi_2
\nonumber\\&&-\frac{\alpha}{2}\frac{(\phi_1-\phi_{cr})}{\phi_{cr}}|{\bf P}|^2+\frac{\alpha}{4}|{\bf P}|^4+\frac{\kappa}{2}(\nabla{\bf P})^2.
\end{eqnarray}
The first two terms, multiplied by the positive constant $a$, guarantee the existence of four coexisting minima, $\phi_1\simeq \phi_0$ and $\phi_2\simeq \phi_0$ representing the active layer and the passive droplet while $\phi_1\simeq 0$ and $\phi_2\simeq 0$ determining the continuous phase.
The third term gauges the energetic penalty due to the fluid interfaces and the fourth one is a repulsive contribution preventing the coalescence of the phase fields. Both $k$ and $\epsilon$ are positive constants. The last three contributions are borrowed from liquid crystal theory and comprise bulk free energy terms multiplied by $\alpha$ and distortions ones multiplied by $\kappa$ \cite{degennes}. Also, $\phi_{cr}=\phi_0/2$ is the critical concentration at which the transition from the isotropic phase (where $|{\bf P}|=0$) to the polar one (where $|{\bf P}|>0$) occurs.

The fields $\phi$ and ${\bf P}$ generally obey advection-diffusion equations, in which the time evolution is governed by thermodynamic forces obtained by functional differentiation of the free energy ${\cal F}=\int fdV$. 

The equation of the scalar phase field $\phi_i$ is 
\begin{equation}\label{eq_phi}
\frac{\partial\phi_i}{\partial t}+{\bf v}\cdot\nabla\phi_i=M\nabla^2\mu_i,
\end{equation}
where $M$ is the mobility and $\mu_i=\delta{\cal F}/\phi_i$ is the chemical potential.

The equation of the polar field ${\bf P}$ is given by
\begin{equation}\label{P_eq}
\frac{\partial {\bf P}}{\partial t}+{\bf v}\cdot\nabla{\bf P}+{\underline {\Omega}}\cdot{\bf P}-\xi {\underline {\mathbf D}}\cdot{\bf P} =-\frac{1}{\Gamma}\frac{\delta {\cal F}}{\delta {\bf P}},
\end{equation}
where the terms on the left hand side represent a generalized material derivative with ${\underline {\Omega}}=({\underline {\mathbf W}}-{\underline {\mathbf W}}^T)/2$,
${\underline {\mathbf D}}=({\underline {\mathbf W}}+{\underline {\mathbf W}}^T)/2$ and $W_{\alpha\beta}=\partial_{\beta}v_{\alpha}$ (Greek indexes denote Cartesian components). 
We set $\xi>1$ as usually done for flow-aligning rodlike particles. Also, $\Gamma$ is the collective diffusion rotational constant multiplying the molecular field ${\bf h}=\delta {\cal F}/\delta {\bf P}$.

The density $\rho$ and the fluid velocity ${\bf v}$ are governed by the Navier-Stokes equations which, in the incompressible limit, are
\begin{equation}\label{cont_eq}
\nabla\cdot{\bf v}=0,
\end{equation}
\begin{equation}\label{nav_stok}
  \rho\left(\frac{\partial}{\partial t}+{\bf v}\cdot\nabla\right){\bf v}=-\nabla p + \nabla\cdot({\underline\sigma}^{\textrm{active}}+{\underline\sigma}^{\textrm{passive}}),
\end{equation}
where $p$ is the isotropic pressure. The stress term on the right hand side is the sum of two contributions, an active and a passive one. The former is given by
\begin{equation}\label{act_str}
\sigma^{active}_{\alpha\beta}=-\zeta\phi_1(P_{\alpha}P_{\beta}-\frac{1}{d}{{\bf P}^2\delta_{\alpha\beta}}),
\end{equation}
where $d$ is the dimension of the system and $\zeta$ is the active parameter, positive for extensile fluids and negative for contractile ones \cite{ramaswamy}. Note that Eq.\ref{act_str} depends on $\phi_1$ since the layer is the sole active component of the system. The functional form of Eq.\ref{act_str} results from a coarse graining over a collection of microscopic force dipoles generated by each extensile unit, which pulls the fluid inwards equatorially and emit it axially.

The passive stress consists of three further contributions, namely a viscous one given by
\begin{equation}
\sigma_{\alpha\beta}^{viscous}=\eta(\partial_{\alpha}v_{\beta}+\partial_{\beta}v_{\alpha})
\end{equation}
where $\eta$ is the shear viscosity, an elastic one given by
\begin{equation}\label{el_st}
\sigma_{\alpha\beta}^{\textrm{elastic}}=\frac{1}{2}(P_{\alpha}h_{\beta}-P_{\beta}h_{\alpha})-\frac{\xi}{2}(P_{\alpha}h_{\beta}+P_{\beta}h_{\alpha})-\kappa\partial_{\alpha}P_{\gamma}\partial_{\beta}P_{\gamma},
\end{equation}
due to the liquid crystal bulk deformations, and an interfacial one 
\begin{eqnarray}\label{int_st}
\sigma_{\alpha\beta}^{\textrm{interface}}&=&\left(f(\phi_1,{\bf P})-\phi_1\frac{\delta{\cal F}}{\delta\phi_1}\right)\delta_{\alpha\beta}-\frac{\partial f(\phi_1,{\bf P})}{\partial(\partial_{\beta}\phi_1)}\partial_{\alpha}\phi_1\nonumber\\&+&
\left(f(\phi_2)-\phi_2\frac{\delta{\cal F}}{\delta\phi_2}\right)\delta_{\alpha\beta}-\frac{\partial f(\phi_2)}{\partial(\partial_{\beta}\phi_2)}\partial_{\alpha}\phi_2.    
\end{eqnarray}

\subsection{Numerical aspects}

Eqs.(\ref{eq_phi})-(\ref{P_eq})-(\ref{cont_eq})-(\ref{nav_stok}) are numerically solved by a hybrid lattice Boltzmann (LB) method \cite{succi,vergassola,higuera,marenduzzo_pre}, where a standard LB approach is used to integrate Eq.\ref{cont_eq} and Eq.\ref{nav_stok} and a finite-difference scheme for  Eq.\ref{eq_phi} and Eq.\ref{P_eq}.  

Our 2D simulations are run on rectangular boxes of size $L_z=140$ (vertical one) and $L_y=400$ (horizontal one) lattice sites. Periodic boundary conditions are set along the $y$ direction while two flat parallel walls are placed at $z=0$ and $z=L_z$. A passive fluid droplet is placed in the middle of the microchannel and a layer of extensile fluid is glued at the bottom wall (see Fig.\ref{fig1}). 
The scalar fields are initialized as follows: $\phi_1\simeq 2$ within a layer of height ranging from $h=10$ to $h=30$ lattice sites (thus occupying a portion from $5$ to $20$ percent of the channel) and zero everywhere else, while $\phi_2\simeq 2$ solely within the passive droplet of radius $R=20$. Also, the polar field is set uniform and parallel to the $y$ direction (i.e. ${\bf P}({\bf r},0)={\bf P}_y({\bf r},0)$ with $ P_y=1$) within the layer and zero outside.
We impose no-slip conditions  at both walls for ${\bf v}$, i.e. $v|_{z=0,L_z}=0$, tangential anchoring at the bottom wall for ${\bf P}$, i.e. $P|_{z=0}=P_y$ and neutral wetting for $\phi_i$, meaning that 
${\bf n}\cdot\nabla\mu_i|_{z=0,L_z}=0$ (no mass flux through the walls) and ${\bf n}\cdot\nabla(\nabla^2\phi_i)|_{z=0,L_z}=0$ (interfaces perpendicular to the walls), with ${\bf n}$ unit vector normal to the boundaries. The addition of an active layer at the top wall follows an analogous prescription, i.e. a further scalar field identifying a region of thickness ranging from $10$ to $30$ lattice sites with a uniform and unidirectional polarization pointing towards the $y$-direction. Accordingly, neutral wetting and strong planar anchoring are set at the wall. 

Following previous works \cite{demagistris,tjhung2,tiribocchi_nat}, thermodynamic parameters have been chosen as follows: $a=4\times 10^{-2}$, $k=6\times 10^{-2}$, $\epsilon=5\times 10^{-3}$, $\alpha=10^{-1}$, $\kappa=4\times 10^{-2}$, $\xi=1.1$, $\Gamma=1$, $M=10^{-1}$, $\eta=1.67$, while $\zeta$ is generally varied between $\simeq 10^{-3}$ and $\simeq 5\times 10^{-2}$. Parameters $a$ and $k$ determine the value of the surface tension $\sigma$ and interface thickness $\lambda$ which, for simplicity, are same for droplet and active layer.  One has $\sigma=\sqrt{8ak/9}\simeq 4.5\times 10^{-2}$ and $\lambda=2\sqrt{2k/a}\simeq 4$. Also, lattice spacing and integration timestep are fixed to $\Delta x=1$ and $\Delta t=1$.
Suitable dimensionless quantities capturing the physics at the mesoscale are the Reynolds number, the Capillary number and the Ericksen number. The first one, defined as $Re=\rho v_{d} D /\eta$ (where $v_{d}$ is droplet speed at the steady state and $D$ its diameter) ranges from $\sim 5\times 10^{-2}$ to $\sim 1$, whereas the second one, defined as $Ca=v_d\eta/\sigma$ varies between  $\sim 4\times 10^{-2}$ and $\sim 7\times 10^{-1}$. Such values ensure that the flow is laminar and droplet breakup is a very unlikely event. 
Finally, the Ericksen number is defined as $Er=\frac{\zeta h_l}{\sigma}$, where $h_l$ is the height of the layer. On a general basis, if this parameter is larger than $1$ the active forces would overcome the resistance to deformation of the liquid crystal (controlled by the elastic constant $\kappa$), thus inducing a spontaneous flow. In our simulations, a sufficiently intense flow capable of triggering the motion of the passive droplet is found at $Er\simeq 1.3$, for $h_l\simeq 30$. Note that this effect occurs only if the polarization (initially uniform everywhere in the layer, with ${\bf P}={\bf P}_y$) is destabilized by a weak perturbation, such as a change of the local orientation of the polar field, lasting for a short period of time. 
Otherwise, in a perfectly ordered and parallel collection of extensile units, the self-generated force dipoles would cancel and no spontaneous flows would be produced.

A mapping to real units can be built considering that a unit of time, length and force in our simulations are approximately equal to $L=1\mu$m, $T=1$ms and $F=100$nN. Thus, our numerical experiment would simulate a droplet of diameter $D\simeq 40\mu$m with $\sigma\simeq 4$mN/m moving at speed ranging from $1$ to $10\mu$m/s (corresponding to  values varying between $10^{-3}$ and $10^{-2}$) in a Newtonian fluid, such as water. The active layer of thickness ranging from $\simeq 10$ to $\simeq 30\mu$m would  consist of an extensile material (such as a bacterial suspension) with a shear viscosity $\eta\simeq 0.1$kPa$\cdot$s, elastic constant $\kappa\simeq 4$nN and activity $\zeta\simeq 10^2$Pa.

\section{Results}

We start off by discussing the dynamics of the passive droplet whose motion is triggered by the active layer glued at the bottom wall of a microfluidic channel. Afterwards we study the effect produced by a second active layer attached at the opposite wall and, finally, we showcase a selection of results obtained for a 3d channel. In all such cases, the drop is initially placed in the middle of the microchannel while the layers uniformly cover the walls. Once drop and layer are relaxed towards equilibrium, the liquid crystal is activated by setting $\zeta$ to values capable of generating a sufficiently high spontaneous flow.

\subsection{Single active layer}\label{sal}

We generally identify a regime for $\zeta\lesssim 4.5\times 10^{-3}$ where the spontaneous flow  triggers a unidirectional motion, while for $\zeta>4.5\times 10^{-3}$ the flow destabilizes and breaks the layer.

\begin{figure*}[htbp]
\includegraphics[width=1.0\linewidth]{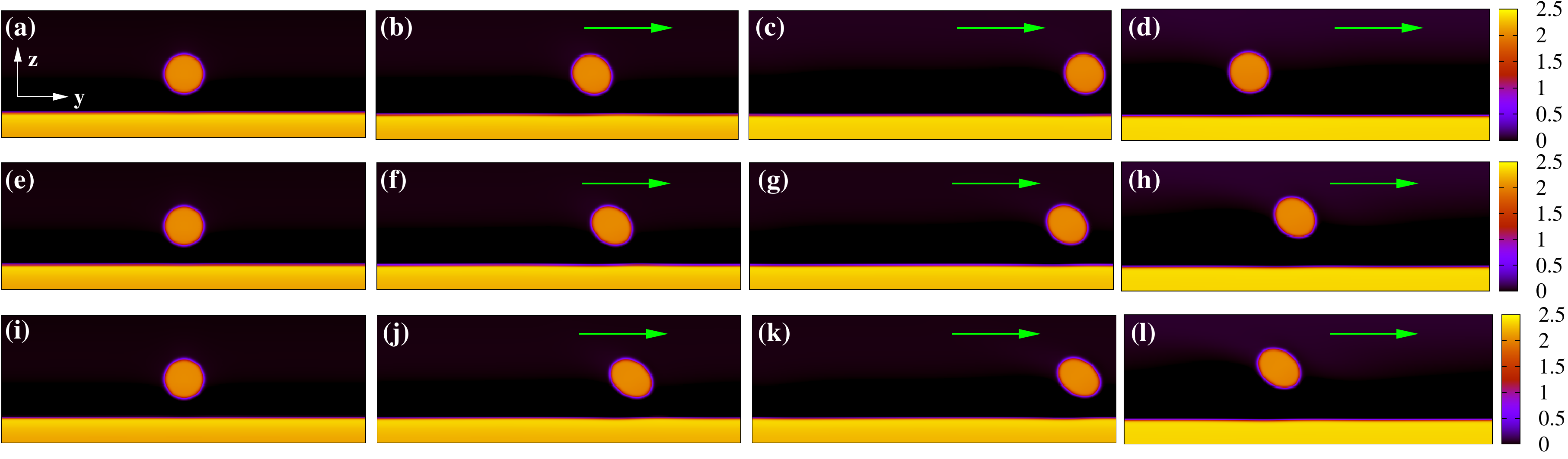}
\caption{Motion of a passive fluid droplet within an active microchannel, in which an extensile material covers the bottom wall. In (a)-(d)  $\zeta=2\times 10^{-3}$, in (e)-(h) $\zeta=3\times 10^{-3}$ and in (i)-(l) $\zeta=4\times 10^{-3}$. In all cases, the unidirectional motion of the droplet is caused by the spontaneous flow generated by the active layer. For increasing values of $\zeta$, droplet speed and shape deformations augment. The green arrows indicate the direction of motion. The color bar represents the values of the phase field of both passive and active fluids.}
\label{fig1}
\end{figure*}

\begin{figure*}[htbp]
\includegraphics[width=1.0\linewidth]{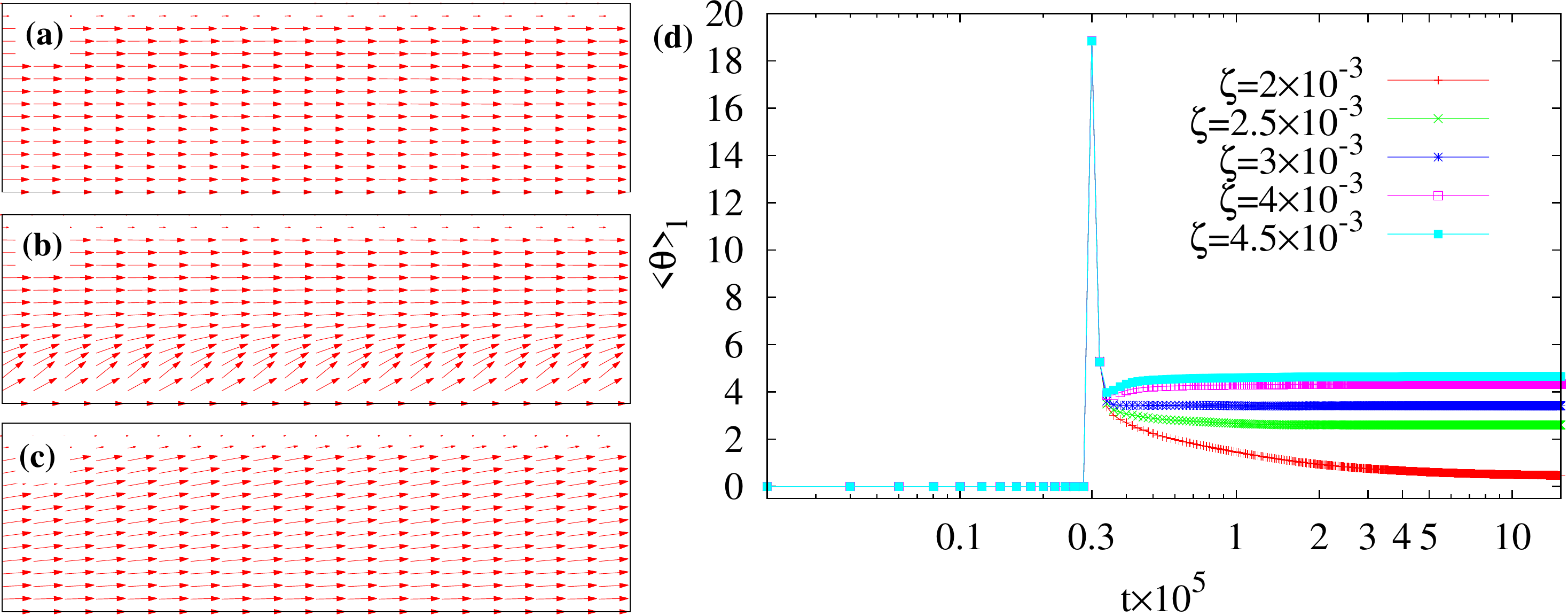}
\caption{Typical polarization profile of a portion of the layer at equilibrium (a), at the onset of the instability (b) and at the steady state (c) for $\zeta=4\times 10^{-3}$. The direction of the polar field, initially uniform and parallel to the $y$ direction (a), slightly tilts near the bottom wall for a short period of time (b), and then gradually relaxes towards a steady orientation (c). The magnitude of {\bf P} is equal to $1$ almost everywhere, except at the interface with the isotropic phase. (d) Time evolution (log scale on the time axis) of the angle $\langle\theta\rangle_l$ of the polarization with respect to the $y$ axis averaged over a layer of thickness $l<h$.}
\label{fig4}
\end{figure*}

In Fig.\ref{fig1} we show a time sequence of the droplet dynamics  for $h/L_z\simeq 0.2$, $R=20$ and $2\times 10^{-3}\lesssim\zeta\lesssim 4.5\times 10^{-3}$, a range of values for which a long-lasting motion can be observed (see also ESI Movie M1, ESI Movie M2 and ESI Movie M3). As previously mentioned, this  is sustained by the flow produced by the active liquid crystal located within the layer. Since the polar field is initially uniform (Fig.\ref{fig4}a), the extensile force pairs of each active unit nullify. Hence, in order to unbalance these forces and generate a spontaneous flow, the orientation of the field above the bottom wall is slightly tilted of an angle $\theta$ (with $0<\theta<\frac{\pi}{2}$), a process that concurrently changes the orientation of the surrounding liquid crystal except that at the walls, which remains fixed (Fig.\ref{fig4}b). This perturbation lasts for a period of time sufficient to destabilize the direction of the polarization, while preserving the integrity of the layer. Later on, the polar field recovers a uniform orientation albeit not parallel to the wall anymore (see Fig.\ref{fig4}c). A rough estimate of its slope can be evaluated by computing the time evolution of the angle $\langle \theta\rangle_l$ averaged over a layer of thickness $l<h$ (with $l\simeq 10$, see Fig.\ref{fig4}d), which identifies a region where a substantial change of orientation occurs.  Once the field is perturbed, $\langle\theta\rangle_l$ shows a sharp peak at approximately $20$ degrees, followed by a rapid relaxation towards a steady state whose value generally increases with $\zeta$. For activities lower than $2\times 10^{-3}$, $\langle\theta\rangle_l$ would approach zero, thus drastically weakening the spontaneous flow.

\begin{figure*}[htbp]
\includegraphics[width=1.0\linewidth]{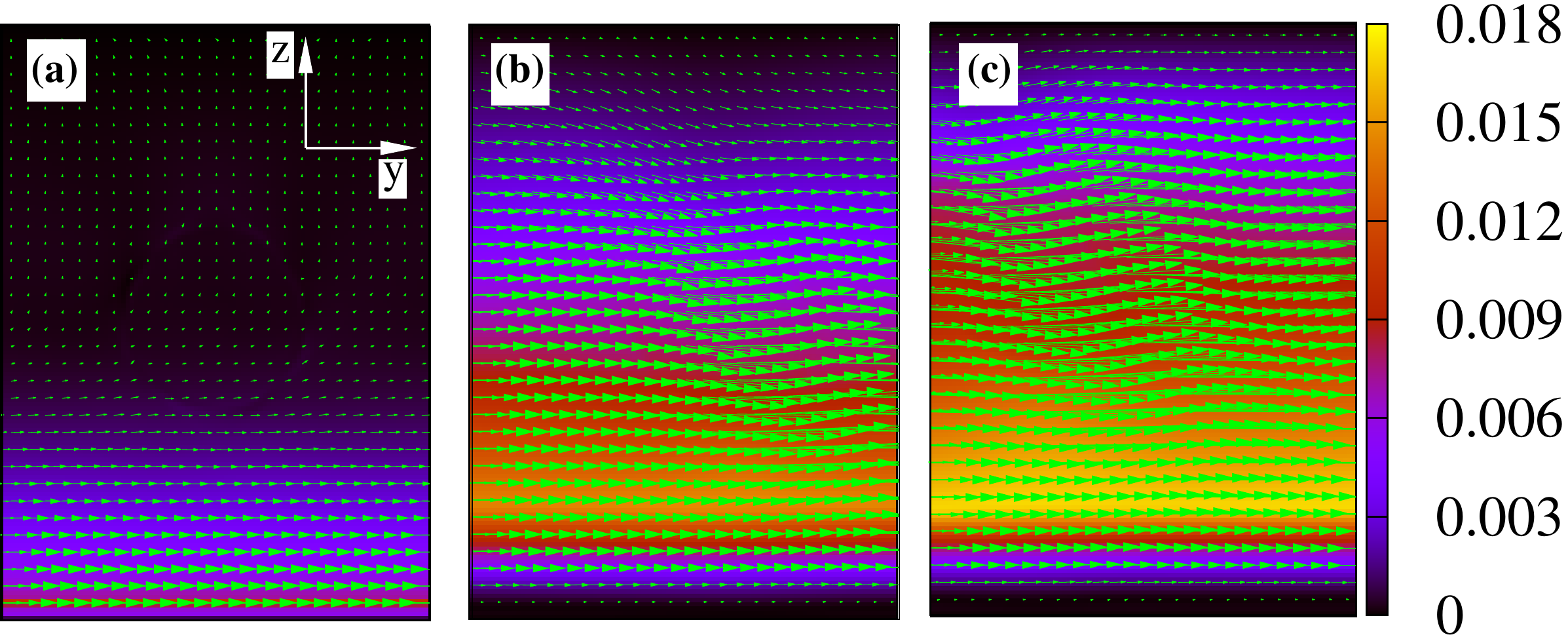}
\caption{Typical velocity field of a portion of lattice in the middle of the microchannel for $\zeta=4\times 10^{-3}$ at the onset of the instability (a-b) and at the steady state (c). It is initially much larger within the layer, and it becomes unidirectional in the whole channel later on. Weak oscillations are caused by the presence of the droplet, which is approximately located in the middle of the channel. The color bar represents the magnitude of ${\bf v}$.}
\label{fig5}
\end{figure*}

\begin{figure*}[htbp]
\includegraphics[width=1.0\linewidth]{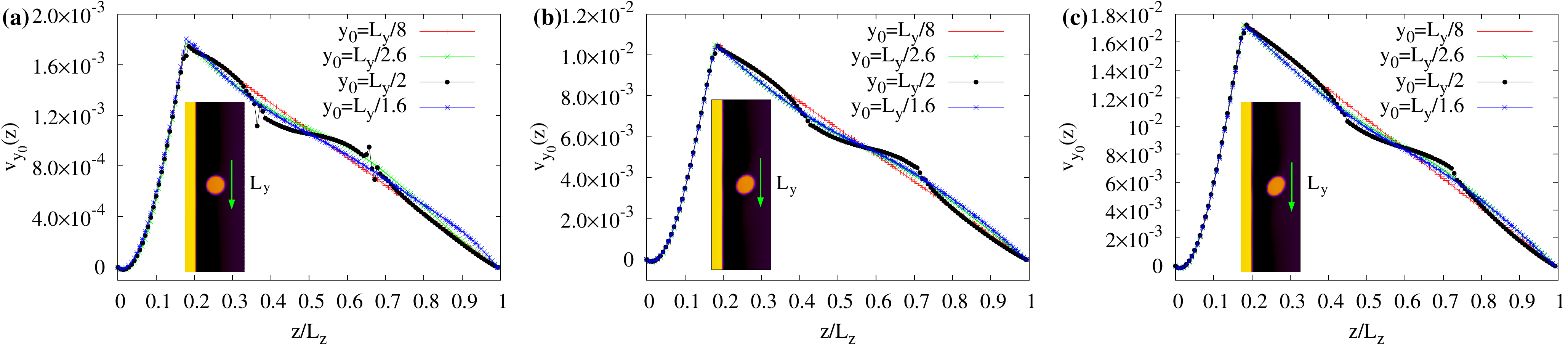}
\caption{(a)-(c). Cross-sectional velocity profiles at different positions in the channel ($y_0=L_y/8,L_y/2.6,L_y/2,L_y/1.6$) computed at the steady state for $\zeta=2\times 10^{-3}$ (a), $\zeta=3\times 10^{-3}$ (b) and $\zeta=4\times 10^{-3}$ (c). The velocity displays a saw-toothed profile, whose maximum is located near the interface of the layer (around $z/L_z\simeq 0.2$) and augments with $\zeta$. Oscillations of the profile observed for $y_0=L_y/2$ are caused by the presence of the droplet. The inset shows the instantaneous position of the droplet within the microfluidic channel and the green arrow indicates the direction of motion.}
\label{fig3}
\end{figure*}

This mechanism produces a velocity field whose typical structure is shown in Fig.\ref{fig5}. At the onset of the instability  the velocity is larger within the layer (Fig.\ref{fig5}a) and, afterwards, it becomes  essentially unidirectional in the whole channel (except in the surrounding of the droplet where weak oscillations form, see Fig.\ref{fig5}b), preserving this pattern for long periods of time (Fig.\ref{fig5}c).
Despite an apparent regularity, a more careful inspection of the velocity field unveils crucial non-uniform features. In Fig.\ref{fig3}a,b,c we plot its cross-sectional profile at different positions along the channel at the steady state, and show that the velocity generally displays a saw-toothed shape, growing linearly within the active layer (for $0<z/L_z\lesssim 0.2$, with a sharp peak at $z/L_z\simeq 0.2$)  essentially because the amount of material increases with the distance from the wall, and decreasing outside towards the opposite wall. Weak undulations of the profile observed at $y_0=L_y/2$ are, once again, due to the presence of the droplet, which is located in the middle of channel. Close to the wall at $z/L_z\simeq 1$, the slope of the curve levels out the one in the bulk whereas, near the opposite wall at $z=0$, it is basically flat because the strong and parallel anchoring of the liquid crystal stabilizes a tiny stress-free region. Indeed, at the steady state one has $\langle\sigma^{tot}_{yz}\rangle_s\sim 0$,
where $\sigma_{yz}^{tot}=\sigma_{yz}^{active}+\sigma_{yz}^{viscous}+\sigma_{yz}^{elastic}+\sigma_{yz}^{interface}$, averaged over a layer of thickness $s$ near the wall. A more detailed description of the effect produced by a change of liquid crystal orientation at the wall is given later, in the paragraph focused on the boundary conditions.

\begin{figure*}[htbp]
\includegraphics[width=1.0\linewidth]{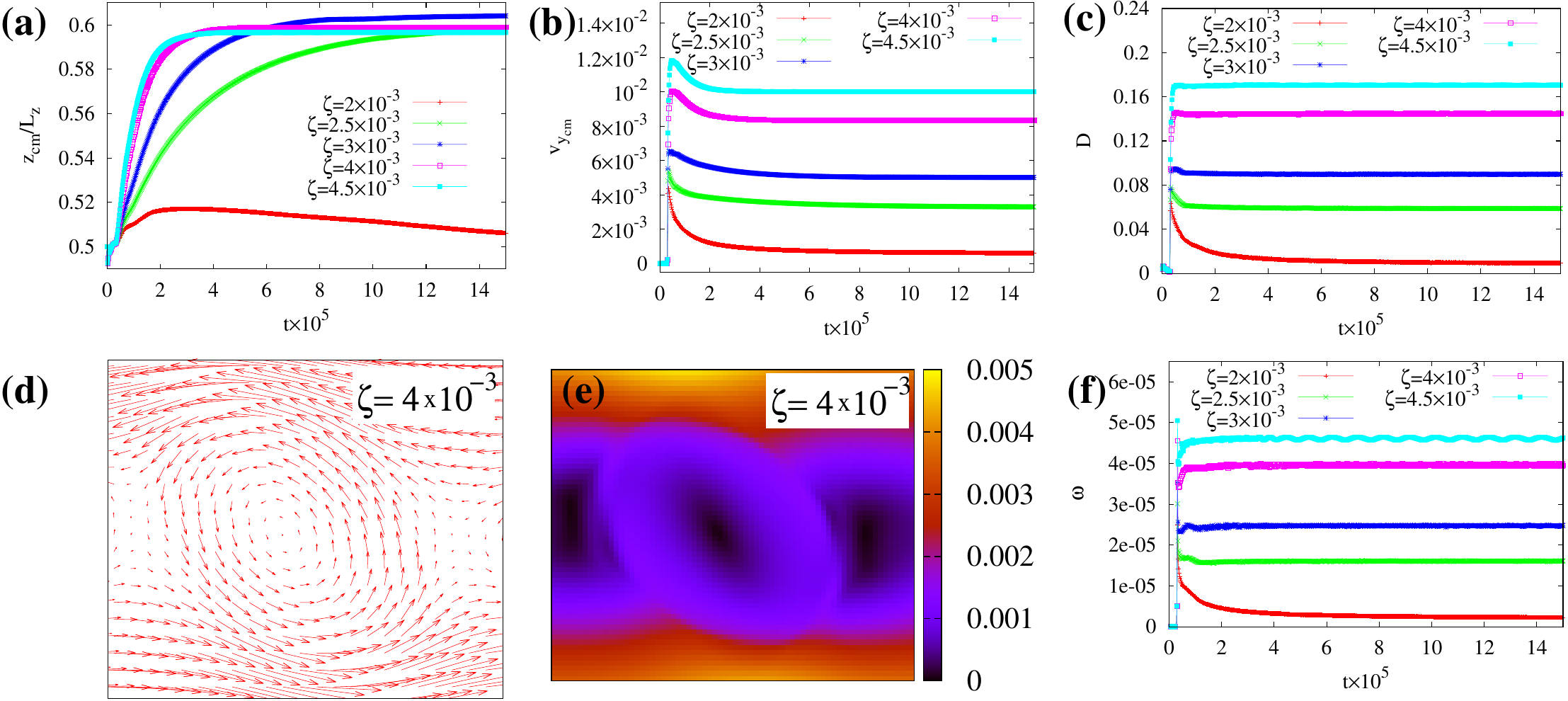}
\caption{(a) Time evolution of $z$-component of droplet center of mass. The drop is shifted towards the upper wall due to the initial asymmetry of the flow, higher within the layer and weaker outside. (b) Time evolution of the $y$-component of the speed of the center of mass.  At early times it displays a peak boosting the drop and, later on, it relaxes to a steady state, whose values increase with $\zeta$. (c) Time evolution of the droplet deformation $D$. Increasing activities trigger larger morphological deformations, turning a circular drop into an elliptical-shaped one.
(d)-(e) Direction and magnitude of the velocity field in the comoving frame of the droplet for $\zeta=4\times 10^{-3}$. (f) Time evolution of the angular velocity of the droplet for different values of $\zeta$. Its behavior is overall akin to that of $v_{y_{cm}}$.}
\label{fig2}
\end{figure*}

As previously discussed, this flow field induces a persistent motion of the droplet along a rectilinear trajectory and essentially at a constant speed. This is shown in Fig.\ref{fig2}a-b where the time evolution of the $z$ component of the center of mass and $y$ component of its speed are plotted for different values of $\zeta$. Once $\zeta$ is turned on, the drop shifts towards the upper wall because of the initial asymmetry of the velocity field (larger in the active layer and weaker outside, see Fig.\ref{fig5}a), an effect that boosts the early time motion (see the peak of $v_{y_{cm}}$). Then, the droplet stabilizes at $z/L_z\simeq 0.6$ for sufficiently high values of $\zeta$ and travels at a constant velocity (whose values increase with $\zeta$). Note that the lift-like effect towards the top wall is also conditioned by the Reynolds and Capillary numbers. Indeed, while for low values of $\zeta$ (thus weak flows) the drop essentially moves along the mid-line of the channel (see Fig.\ref{fig2}a, $\zeta=2\times 10^{-3}$), for higher values it is subject to more intense flows and large shape deformations, which favour the shift upwards (see Fig.\ref{fig1}h,l and Fig.\ref{fig2}a).

Such morphological changes are quantitatively captured by the Taylor parameter $D=\frac{a-b}{a+b}$ (where $a$ and $b$ represent the major and minor axis of the droplet), which ranges between $0$ (no deformation, circular drop) to $1$  (maximal deformation, needle-like shape). At early times $D$ generally exhibits a peak becoming milder as $\zeta$ increases, an effect mirroring the one seen for $v_{y_{cm}}$ and  indicating that the drop undergoes a quick deformation once the activity is turned on. Afterwards, $D$ relaxes to a steady state value faster for higher $\zeta$, since a strong flow stabilizes the shape changes easier than a weaker one (Fig.\ref{fig2}c). 

It is worth noting that the motion along a rectilinear trajectory is partially akin to that observed in a microchannel when the drop is subject to a Poiseuille flow (where no-slip conditions hold and the walls are at rest)\cite{coupier}. A well-known feature of this process is that, while in the lab frame the fluid velocity displays a parabolic-like profile, in the comoving frame it would exhibit a couple of counter-rotating vortices resulting from the sheared structure of the flow. Thus, it is natural to ask whether similar structures emerge in our system. 
In Fig.\ref{fig2}d-e we show the typical steady-state velocity ${\bf v}_s$ computed in the droplet frame (i.e. ${\bf v}_s={\bf v}-{\bf v}_{cm}$) for $\zeta=4\times 10^{-3}$.  It exhibits a well-defined counter-rotating vortex with magnitude increasing up to $\simeq 10^{-3}$ towards the edge of the drop. As for $v_{cm}$, the angular velocity is also found to gradually augment for increasing values of $\zeta$. This is shown in Fig.\ref{fig2}f, where we plot the time evolution of $\boldsymbol{\omega}=\frac{\int dV (\phi_2/R^2){\bf r}\times{\bf v}}{\int dV\phi_2}$ for different values of $\zeta$. If, for example, $\zeta=4\times 10^{-3}$, one has $\omega\simeq 4\times 10^{-5}$ and, for a droplet of radius $R\simeq 20$, $v=\omega R\simeq 10^{-3}$, thus a rolling-like condition would be satisfied. This is in contrast with the results of a spinning droplet containing an active spiral-like gel (where the rotation speed is expected to follow a logarithmic-like behavior \cite{kruse,julicher}) and suggests that, although the motion is triggered by the same medium (here placed outside), the drop behaves as a fully passive object.

These results show that a rectilinear motion of the droplet can be generated by a properly designed layer of active material of predefined thickness. 
But what if the thickness decreases?  In ESI Movie M4 we show the dynamics in a microchannel with an active layer of thickness $h=10$ lattice sites ($h/L_z\simeq 0.07$, three times narrower than the previous case) and $\zeta=2.5\times 10^{-2}$. The motion shares features akin to the ones observed for higher values of $h$, although the reduced amount of active material requires higher values of $\zeta$  to set a sufficiently intense spontaneous flow. 

One may finally wonder whether the shape deformations can be mitigated even for droplet moving at high speed (such as for $\zeta=4\times 10^{-3}$). This can be achieved by covering the upper wall with a further active layer, whose design follows that of the opposite one. In the next section we describe the dynamics of the drop precisely in this configuration.

\subsection{Double active layer}

\begin{figure*}[htbp]
\includegraphics[width=1.0\linewidth]{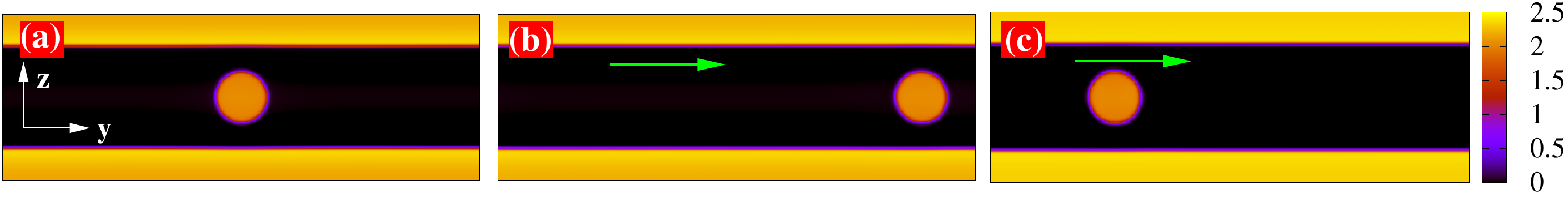}
\caption{Time sequence of the motion of a passive fluid droplet within an active microchannel, in which the extensile material covers bottom and top walls. Here $\zeta=4\times 10^{-3}$. After attaining equilibrium (a), the droplet is set into motion by the spontaneous flow generated by the active layers (b,c). Note that the trajectory remains rectilinear but, unlike the single layer case, shape deformations become negligible. The green arrows indicates the direction of motion. The color bar represents the values of the phase field of both passive and active fluids.}
\label{fig6}
\end{figure*}

In Fig.\ref{fig6} we show a time sequence of a passive droplet moving within a microchannel with two symmetric layers for $\zeta=4\times 10^{-3}$ (equal in both layers). As described in the section Methods, within the upper region one has $\phi_2\simeq \phi_0$ and $\phi_1\simeq 0$, while the polarization is initially parallel to the wall and points along the positive $y$ direction. Also, the thickness of both layers is $h=30$. The mechanism triggering the motion essentially mirrors the one discussed for the single layer.  Once the orientation of the polar field is weakly titled of an angle $\theta_b$ (with $0<\theta_b<\frac{\pi}{2}$ with respect to the $y$ direction) from the bottom wall and $\theta_t$ from the top one (with $\theta_t=2\pi-\theta_b$), the activity is turned on and a spontaneous flow emerges. 

\begin{figure*}[htbp]
\includegraphics[width=1.0\linewidth]{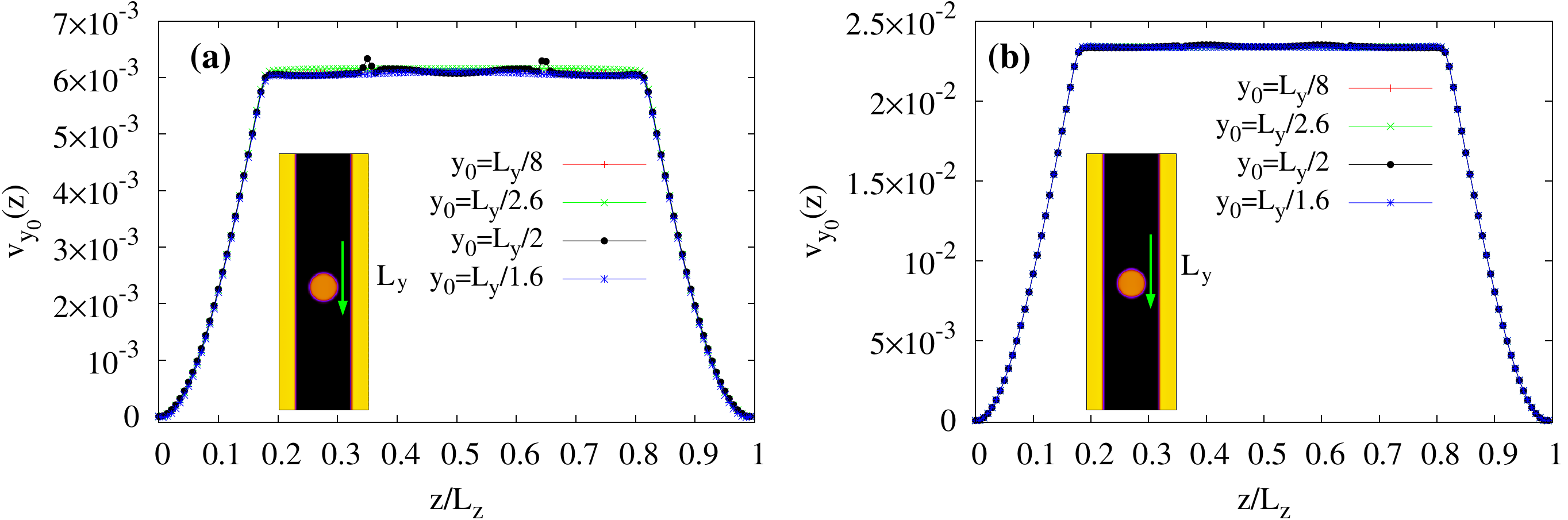}
\caption{Cross-sectional velocity profiles at different positions in the channel ($y_0=L_y/8,L_y/2.6,L_y/2,L_y/1.6$) computed at the steady state for $\zeta=2\times 10^{-3}$ (a) and $\zeta=4\times 10^{-3}$ (b). The velocity augments going from the walls to the interface of the layer and remains uniform within the passive channel. Profile fluctuations, due to the presence of the droplet, become milder as $\zeta$ increases.}
\label{fig8}
\end{figure*}

Unlike the previous case, here the fluid velocity displays a symmetric structure, uniform within the passive fluid (in the middle of the channel, black region of Fig.\ref{fig6}) and growing almost linearly from the wall towards the interfaces of each layer (the yellow regions of Fig.\ref{fig6}) with equal slope. This is shown in Fig.\ref{fig8}, where we plot the steady state cross sectional profile, at different positions along the channel, for various values of $\zeta$.  For increasing activity, $v_{y_0}$ attains higher values and the fluctuations due to the drop interface become progressively negligible. In analogy to the single layer case, the slope of the speed near both walls is approximately flat, an effect caused by the orientation of the liquid crystal (see the section on the boundary conditions).

\begin{figure*}[htbp]
\includegraphics[width=1.0\linewidth]{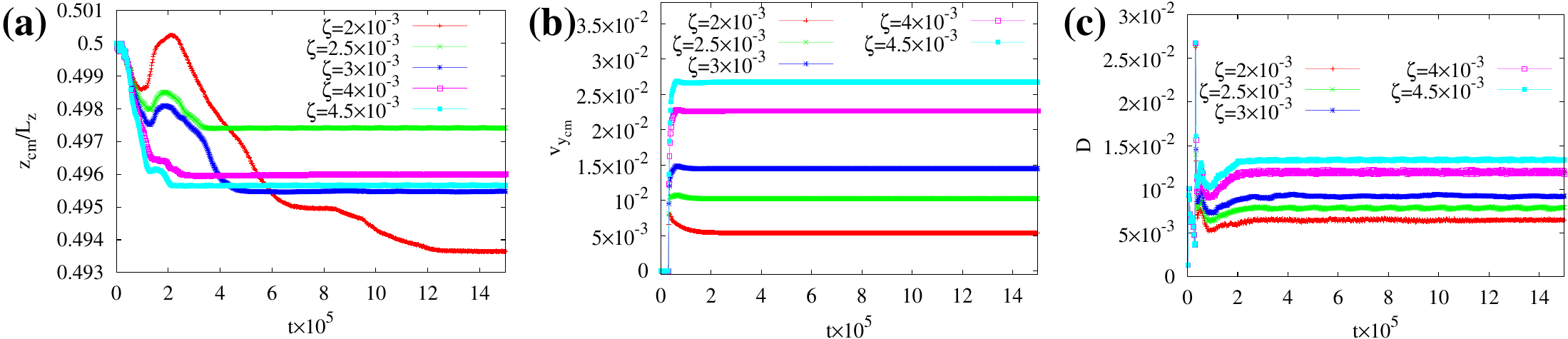}
\caption{(a)-(b) Time evolution of $z$-component of droplet center of mass and $y$-component of its speed. Once the activity is turned on, the drop moves at constant speed along the mid-line. (c) Time evolution of the deformation parameter $D$. The addition of the layer essentially suppresses shape deformations.}
\label{fig7}
\end{figure*}

This flow triggers, once again, a rectilinear motion at constant speed,  keeping the droplet along the mid-line of the channel, regardless of the values of $\zeta$ (see Fig.\ref{fig7}a,b). In addition, flat pattern of the velocity profile in the passive fluid suppresses morphological deformations (see Fig.\ref{fig7}c) and stabilizes a droplet that keeps a long-lasting circular shape. Note also that the speed of the drop is approximately twice than that obtained in the single layer, provided that $\zeta$ is set to equal values in both regions. 

Interestingly, the flow structure closely resembles the electro-osmotic flow observed across porous materials and capillary tubes\cite{yang,tao,kundu}. In this case, the flow is caused by an electric field (applied, for instance, via electrodes) which induces the motion of a net charge forming two tiny layers of mobile ions, sandwiched between the  wall of the microchannel and the bulk electrolyte solution. The resulting flow field is found to display a growing profile within the electrical layers (which would correspond to the yellow layers of Fig.\ref{fig6}) while  a uniform, flat one, in the intermediate solutions (which would correspond to the black region of Fig.\ref{fig6}).

\subsection{Shear-like dynamics}

The dynamics described so far ultimately results from the orientation of the polar field which, in both layers, points towards the positive $y$ axis forming an angle $\theta_b$ with the bottom wall and $\theta_t=2\pi-\theta_b$ with the top one. An alternative scenario emerges if the polarization is oriented along opposite directions in the layers.

\begin{figure*}[htbp]
\includegraphics[width=1.0\linewidth]{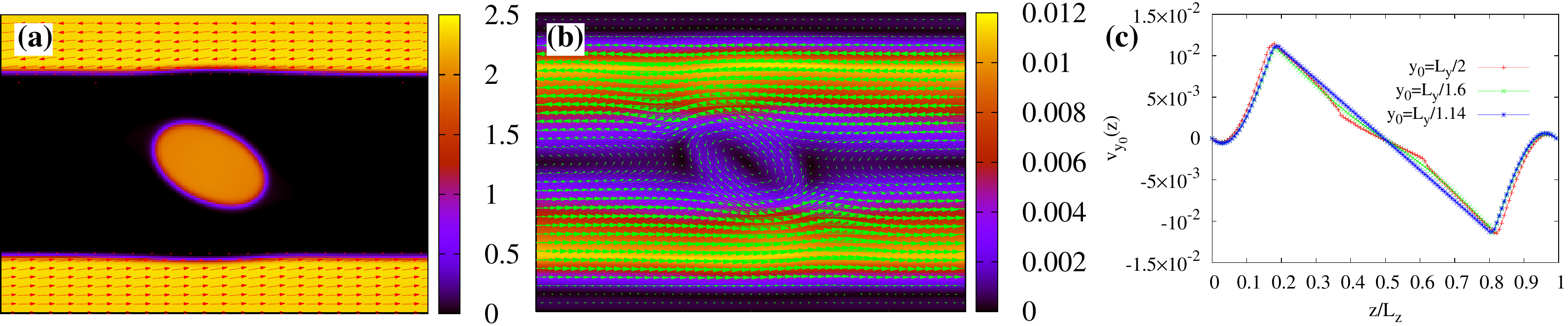}
\caption{(a) Steady states of droplet and polar field in a portion of the microchannel, observed when the liquid crystal is oriented along opposite directions in the two layers. (b) Corresponding structure of the velocity field. The flow exhibits a bidirectional pattern near the fluid interfaces of the layers plus a counterclockwise vortex in the droplet. (c) Cross-sectional velocity profile at different positions in the channel. Fluctuations in the passive region are due, once again, to the droplet interface.}
\label{fig12}
\end{figure*}

In Fig.\ref{fig12}a we show, for example, the steady state configurations of droplet and polarization under this condition for $\zeta=4\times 10^{-3}$, where the liquid crystal has been initially aligned along the negative $y$ direction in the top layer. While at late times the latter displays the usual uniform arrangement, the droplet attains a non-motile elliptical shape induced by the shear-like structure of the velocity field (shown in Fig.\ref{fig12}b). Indeed, the velocity exhibits a striped pattern, where  an intense  bidirectional flow near the fluid interfaces of the layers combine with a much weaker counterclockwise vortex within the droplet. Near both walls, the flow is once again very low. The corresponding transversal profile is shown in Fig\ref{fig12}c; it is shear-like in the passive region and almost linearly growing in the active layers, with features akin to those discussed in the previous section. Also, near the top wall ($z/L_z\simeq 1$), $v_{y_0}$ has a local maximum necessary to accommodate the stress-free region due to the parallel anchoring of ${\bf P}$.

It is worth highlighting that the stability of this configuration crucially depends on the orientation of the liquid crystal. More specifically, a shear-like dynamics occurs when $\theta_t\simeq\theta_b+\pi$ (with $0<\theta_b<\frac{\pi}{2}$), otherwise the droplet, still keeping an elliptical shape, would move, following a rectilinear trajectory, along the direction imposed by the larger flow, thus rightwards if $\pi<\theta_t<\theta_b+\pi$ (the bottom flow dominates) or leftwards if $\theta_b+\pi<\theta_t<\frac{3}{2}\pi$ (the top one prevails).

\subsection{Boundary conditions}

\begin{figure*}[htbp]
\includegraphics[width=1.0\linewidth]{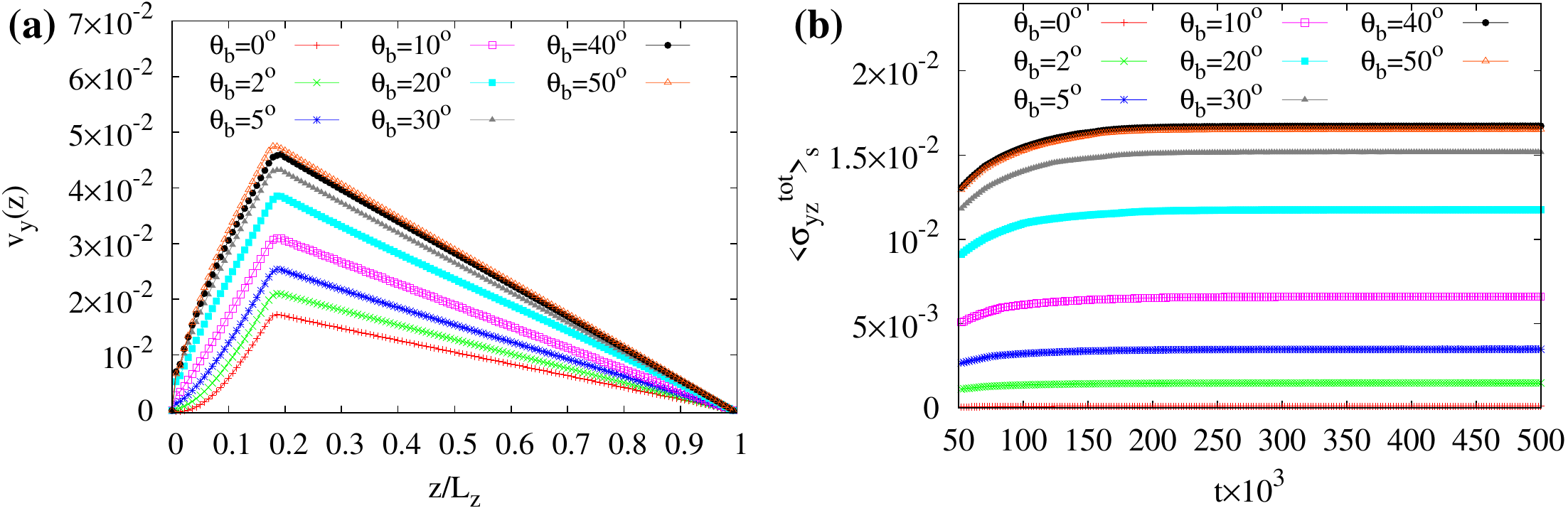}
\caption{(a) Cross sectional profile of the $y$ component of the center of mass speed at the steady state and for different values of anchoring angle $\theta_b$ of the liquid crystal at the bottom wall. The activity is fixed at $\zeta=4\times 10^{-3}$. For increasing values of $\theta_b$, the slope of $v_y$ grows near $z\simeq 0$ and attains a maximum approximately at $\theta_b\simeq 50^o$. (b) Time evolution of total stress $\langle\sigma_{yz}^{tot}\rangle_s=\langle\sigma_{yz}^{active}+\sigma_{yz}^{viscous}+\sigma_{yz}^{elastic}+\sigma_{yz}^{interface}\rangle_s$, averaged over a film of thickness $s=4$ lattice sites.}
\label{fig11}
\end{figure*}

In Fig.\ref{fig3}, Fig.\ref{fig8} and Fig.\ref{fig12}c we have shown that the velocity profile grows almost linearly within active layer and is approximately flat near the wall covered by the active material. The latter crucially depends on the orientation of the liquid crystal at the wall. In Fig.\ref{fig11}a we plot the cross-sectional velocity profile for different anchoring angles $\theta_b$  at the bottom wall for $\zeta=4\times 10^{-3}$. For increasing values of $\theta_b$, the slope of $v_y$ sharpens (thus removing the flatness close to the wall) and the total speed concurrently grows, attaining the highest value at $\theta_b\simeq 50^{\circ}$. Indeed, while for $\theta_b\simeq 0$ the force dipoles (aligned along the vectors ${\bf P}$) nullify each other, for $\theta_b > 0$ 
the balance is broken and an additional spontaneous flow is produced by the active liquid crystal anchored at the boundary. Changing of the anchoring orientation modifies
the total stress too, as shown in  Fig.\ref{fig11}b where we plot the time evolution of $\langle\sigma_{yz}^{tot}\rangle_s$ averaged over a film of thickness $s=4$ lattice sites from the bottom wall.  The stress remains approximately zero for $\theta_b=0$ while it considerably augments for higher values. 

These results suggest that, at fixed activity, changing the tilt of the active gel at the boundary can control the structure of the spontaneous flow and, more generally, could potentially trigger the transition from a passive quiescent state towards and active one. Note that these findings extend the ones of Ref.\cite{mare_yeom}, which describe the formation of spontaneous flows in one dimensional liquid crystal cells only for two different anchoring conditions (perpendicular and conflicting).

\subsection{Extension to 3D systems}

\begin{figure*}[htbp]
\includegraphics[width=1.0\linewidth]{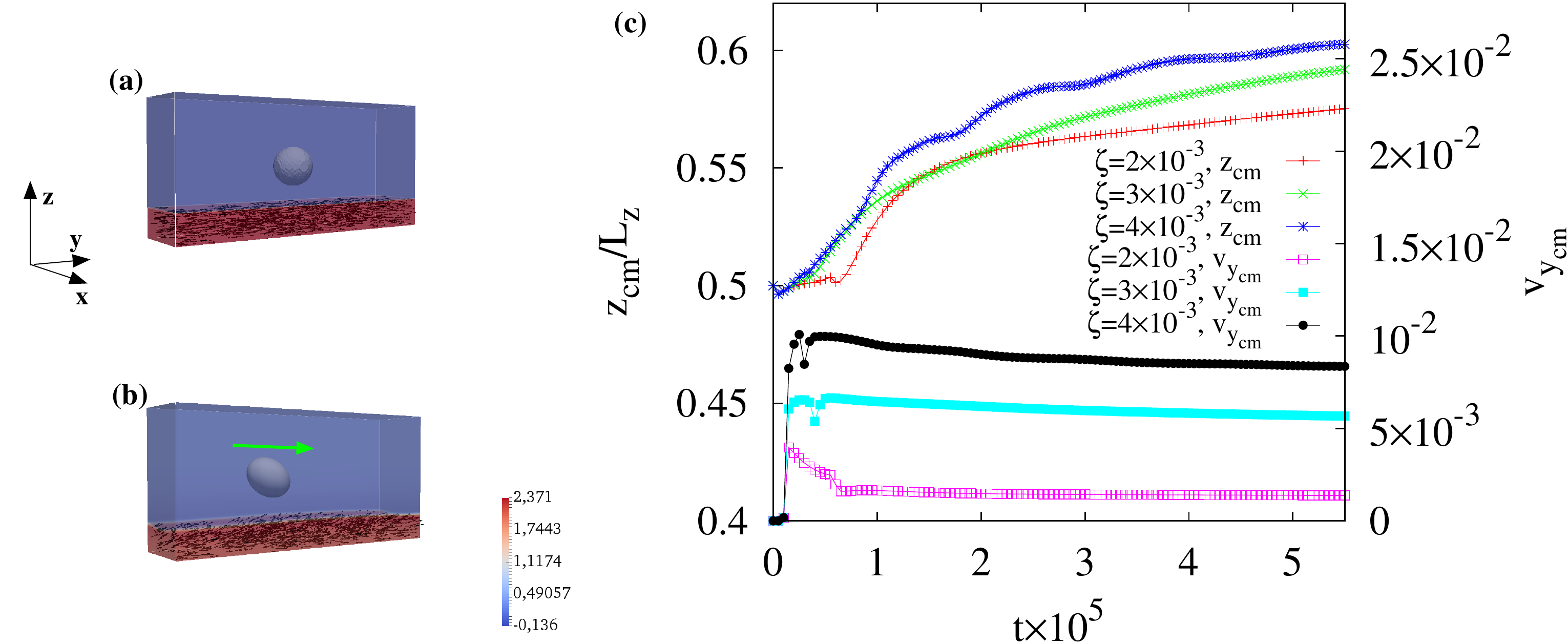}
\caption{Three dimensional simulation of a passive fluid droplet moving within an active microchannel for $\zeta=4\times 10^{-3}$. (a) The droplet is initially surrounded by a passive fluid (blue) in the middle of the channel while a liquid crystal layer (red) covers the bottom wall. Once the activity is turned on, a spontaneous flow triggers the motion of the passive droplet. (b) At the steady state, the droplet acquires a permanent ellipsoidal shape moving along a rectilinear trajectory. The green arrow indicates the direction of motion. The color map shows the range of values of the phase field associated to the active layer. (c) Time evolution of $z$ component of center of mass (left axis) and  $y$ component of its speed (right axis). At high $\zeta$ the trajectory exhibits oscillations (for $10^5\lesssim t\lesssim 4\times 10^5$) triggered by the ones of polar field and fluid velocity.}
\label{fig9}
\end{figure*}

The dynamics found in 2D can be observed in a 3D microfludic channel as well. For simplicity, we consider the case where a single active layer covers the bottom wall and the polarization is initially uniform and parallel to the $y$ axis (see Fig.\ref{fig9}a). The system consists of a passive fluid droplet of radius $R=20$ sandwiched between two flat parallel walls placed at $L_z=0$ and $L_z=140$. Along the $x$ and $y$ directions, periodic boundary conditions are set with $L_x = 70$ and $L_y = 300$. Once the droplet is equilibrated, the activity is turned on and, following the mechanism discussed for the 2D system, a spontaneous flow is produced within the channel. This one, in turn, triggers a rectilinear motion of the droplet which, at the steady state, attains a permanent elliptical shape (Fig.\ref{fig9}b,c and ESI Movie M5). 

\begin{figure*}[htbp]
\includegraphics[width=1.0\linewidth]{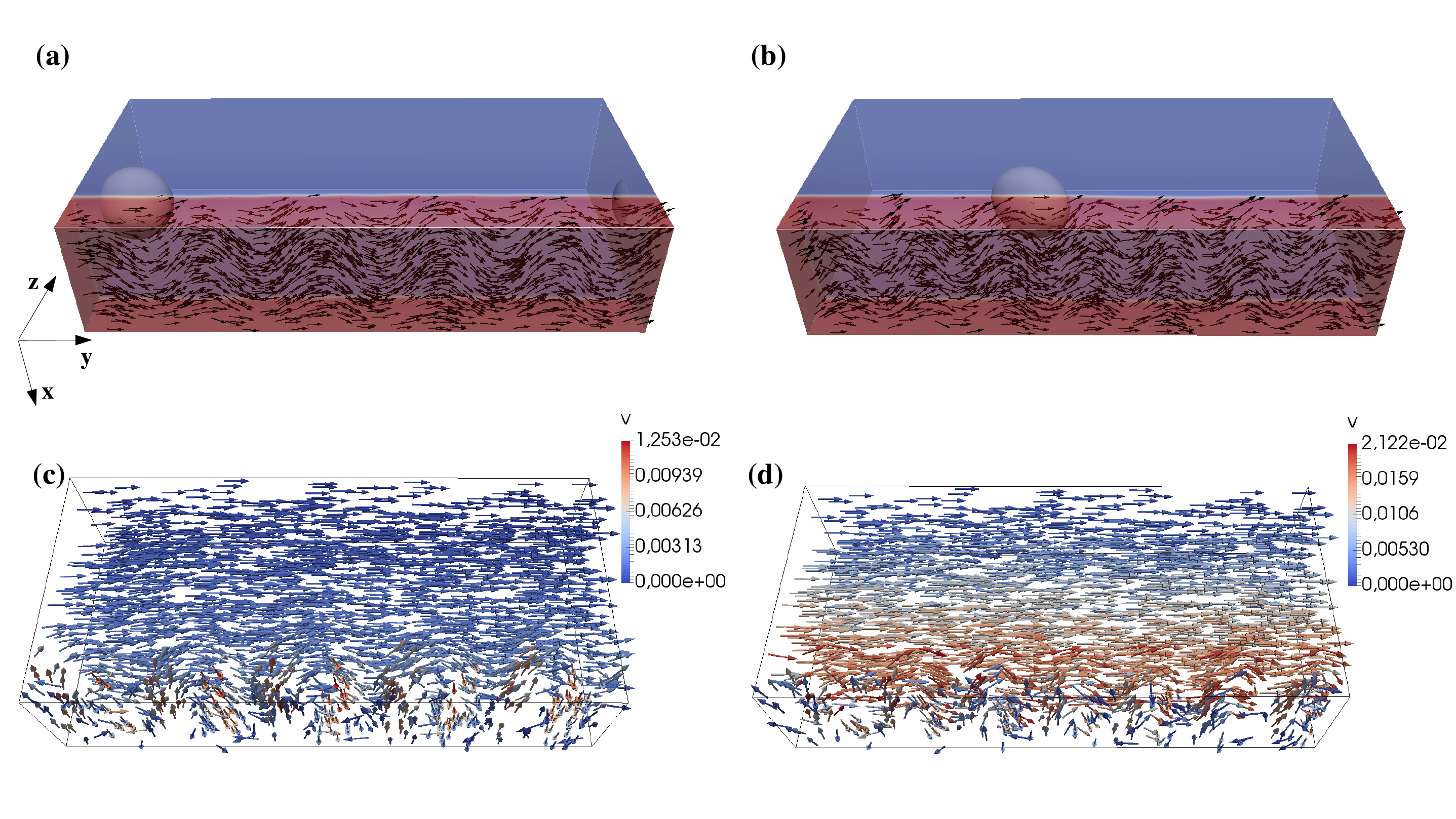}
\caption{(a)-(b) Instantaneous configurations of the polarization within the layer for $\zeta=2\times 10^{-3}$ (a) and $\zeta=4\times 10^{-3}$ (b). At the steady state, the liquid crystal displays pristine in-plane bend distortions, whose number increases for higher $\zeta$.
(c)-(d) The plots show direction and magnitude of the velocity field in the channel. While in the passive fluid the flow is unidirectional along the positive $y$ axis, in the active layer it also shows an undulating pattern in the $xy$ plane. Its direction is locally perpendicular to that of the liquid crystal.}
\label{fig10}
\end{figure*}

Although these results are overall akin to the 2D system, crucial differences emerge in the polarization and velocity field within the layer.
Indeed, the liquid crystal exhibits stable bend deformations mostly in the $xy$ plane, an effect stemming from the hydrodynamic instability produced by dipolar forces of extensile materials made of rod-shaped particles (Fig.\ref{fig10}a,b), while the velocity field displays two distinct patterns: a unidirectional flow within the passive fluid and an undulating motif, with vectors (once again mostly in the $xy$ plane) pointing perpendicularly to the liquid crystal orientation, in the layer (Fig.\ref{fig10}c,d), a result in agreement with the typical flow structure observed in extensile gels \cite{sriram}. It is interesting to highlight that, despite the higher complex pattern of the velocity with respect to the 2D device, the unidirectional motion of the droplet is preserved in 3D (see Fig.\ref{fig9}c). This is precisely because the distortions of the polar field are confined  within the layer (thus the spontaneous flow displays the undulations only within this region) while out of it the fluid is passive and isotropic, and the velocity follows the direction imposed by the one in the layer for continuity. However, at high $\zeta$ (such as $\zeta=4\times 10^{-3}$), the droplet trajectory  and the fluid interface are found to exhibit weak oscillations,  caused by temporary out-of-plane components (along $z$) of the polar field and of the fluid velocity (see Fig.\ref{fig9}c and ESI Movie M5).

\section{Conclusions}

To summarize, we have theoretically studied the dynamics of a passive fluid droplet confined within a microfluidic channel whose walls are covered with a layer of an extensile active gel. The physics is mostly controlled by the activity and the orientation of the liquid crystal both in the layer and at the walls. Our results show that, for sufficiently high values of activity, the passive droplet acquires a rectilinear motion lasting over long periods of time caused by the spontaneous flow produced by the active material. If the active gel covers only one wall, at the steady state the droplet attains an ellipsoidal shape, whose rate of deformation increases with the activity. This is due to the asymmetric structure of the spontaneous flow, which exhibits a saw-toothed pattern with a maximum at the interface of the layer. If, on the contrary, the active material covers both walls, the flow displays a trapezoidal-like symmetric structure and shape deformations turn negligible. In addition, if the liquid crystal points along opposite directions in the layers, a shear-like profile of the flow is found. These behaviors are also affected by the orientation of the liquid crystal at the wall. Indeed, our simulations suggest that, for a fixed value of activity, the magnitude of the flow (and thus the speed of the drop) increases when the anchoring angle at the wall augments. These findings are confirmed by 3D simulations, although liquid crystal and velocity field display a more complex structure especially within the active layer, where an undulating flow favours the formation of well-defined bend distortions. 

In this paper we devise a strategy to control and rectify the trajectory of passive fluid drops confined in an "active" microchannel, in which the necessary work to sustain the motion is extracted from a layer of active material. This result could be in principle extended to active droplets, where maintaining a certain direction of motion remains challenging. Indeed, although an enhanced Brownian motion has been observed in drops containing either a dispersion of bacteria \cite{soto_softmatter,soto_prl} or a suspension of motorized microtubules \cite{dogic}, only recently a controllable propulsion has been realized in active droplets hosted in a nematic environment \cite{lavrentovich} or activated through a Marangoni-like effect \cite{maass}.

Our device could be concretely realized by dispersing an extensile fluid (such as microtubule bundles and kinesin or bacterial suspensions) into a water-oil solution, in which the aqueous medium containing the active material wets the walls. To enforce a specific anchoring, the active fluid could be pinned following the protocols adopted for passive liquid crystals, i.e. through mechanical or chemical treatment of the surface \cite{degennes,abbott}. 
Although not discussed in the present study, it would be relevant to investigate how a predetermined anchoring at the fluid interface (where the active gel may either orient tangentially \cite{linek,guanglai_pre} or perpendicularly \cite{sci_rep_bon}) would affect the motion of the droplet and the structure of the flow.

Finally, it is interesting to discuss some directions of future works. Releasing, for example, the approximation of single viscosity of both fluid components could be a further route to control shape deformations as well as simulate highly viscous drops closely behaving as rigid particles. In this context, the system studied in this work would partially capture the dynamics of a spherical non-deformable particle when the activity is low (as in Fig.\ref{fig1}a-d) whereas for non-spherical solid objects, such as elliptical ones, the dynamics is expected be more complex since in-plane or out-of-plane rotations could be observed \cite{masaeli}. Likewise, abandoning the restriction of small Capillary numbers would be useful to explore regimes in which breakup events are more likely. While this physics is well understood in passive systems (such as a droplet subject to shear flows \cite{stone3,bentley}), less is known when the rupture would be determined by a spontaneous flow, as well as to what extent such flow would impact on critical Capillary numbers, beyond which no-steady state droplet exists.  Our device would also be highly suitable for exploring more complex designs, such as those in which the activity is spatially dependent. One may envisage, for example, a microfluidic channel in which the wall is patterned with separated holes (realized using static phase fields \cite{tiribocchi_nat}) containing an active gel which could trigger a hop-like motion of a drop dispersed in the surrounding passive fluid. This may hold an interest for the manufacturing of active energy-saving devices whose functioning relies on reduced amounts of active material.

\section*{Conflicts of interest}
There are no conflicts to declare.

\section*{Acknowledgements}

The authors acknowledge funding from the European Research Council under the European Union's Horizon 2020 Framework Program (No. FP/2014-2020) ERC Grant Agreement No. 739964 (COPMAT) and ERC-PoC2 grant No. 101081171 (DropTrack). The authors also warmly thank Prof. Ramin Golestanian for useful discussions.

%\balance

%%%REFERENCES%%%
\bibliography{biblio} 

%merlin.mbs apsrev4-1.bst 2010-07-25 4.21a (PWD, AO, DPC) hacked
%Control: key (0)
%Control: author (0) dotless jnrlst
%Control: editor formatted (1) identically to author
%Control: production of article title (0) allowed
%Control: page (1) range
%Control: year (0) verbatim
%Control: production of eprint (0) enabled
\begin{thebibliography}{77}%
\makeatletter
\providecommand \@ifxundefined [1]{%
 \@ifx{#1\undefined}
}%
\providecommand \@ifnum [1]{%
 \ifnum #1\expandafter \@firstoftwo
 \else \expandafter \@secondoftwo
 \fi
}%
\providecommand \@ifx [1]{%
 \ifx #1\expandafter \@firstoftwo
 \else \expandafter \@secondoftwo
 \fi
}%
\providecommand \natexlab [1]{#1}%
\providecommand \enquote  [1]{``#1''}%
\providecommand \bibnamefont  [1]{#1}%
\providecommand \bibfnamefont [1]{#1}%
\providecommand \citenamefont [1]{#1}%
\providecommand \href@noop [0]{\@secondoftwo}%
\providecommand \href [0]{\begingroup \@sanitize@url \@href}%
\providecommand \@href[1]{\@@startlink{#1}\@@href}%
\providecommand \@@href[1]{\endgroup#1\@@endlink}%
\providecommand \@sanitize@url [0]{\catcode `\\12\catcode `\$12\catcode
  `\&12\catcode `\#12\catcode `\^12\catcode `\_12\catcode `\%12\relax}%
\providecommand \@@startlink[1]{}%
\providecommand \@@endlink[0]{}%
\providecommand \url  [0]{\begingroup\@sanitize@url \@url }%
\providecommand \@url [1]{\endgroup\@href {#1}{\urlprefix }}%
\providecommand \urlprefix  [0]{URL }%
\providecommand \Eprint [0]{\href }%
\providecommand \doibase [0]{http://dx.doi.org/}%
\providecommand \selectlanguage [0]{\@gobble}%
\providecommand \bibinfo  [0]{\@secondoftwo}%
\providecommand \bibfield  [0]{\@secondoftwo}%
\providecommand \translation [1]{[#1]}%
\providecommand \BibitemOpen [0]{}%
\providecommand \bibitemStop [0]{}%
\providecommand \bibitemNoStop [0]{.\EOS\space}%
\providecommand \EOS [0]{\spacefactor3000\relax}%
\providecommand \BibitemShut  [1]{\csname bibitem#1\endcsname}%
\let\auto@bib@innerbib\@empty
%</preamble>
\bibitem [{\citenamefont {Marchetti}\ \emph {et~al.}(2013)\citenamefont
  {Marchetti}, \citenamefont {Joanny}, \citenamefont {Ramaswamy}, \citenamefont
  {Liverpool}, \citenamefont {Prost}, \citenamefont {Rao},\ and\ \citenamefont
  {Aditi~Simha}}]{marchetti}%
  \BibitemOpen
  \bibfield  {author} {\bibinfo {author} {\bibfnamefont {M.~C.}\ \bibnamefont
  {Marchetti}}, \bibinfo {author} {\bibfnamefont {J.~F.}\ \bibnamefont
  {Joanny}}, \bibinfo {author} {\bibfnamefont {T.}~\bibnamefont {Ramaswamy}},
  \bibinfo {author} {\bibfnamefont {T.~B.}\ \bibnamefont {Liverpool}}, \bibinfo
  {author} {\bibfnamefont {J.}~\bibnamefont {Prost}}, \bibinfo {author}
  {\bibfnamefont {M.}~\bibnamefont {Rao}}, \ and\ \bibinfo {author}
  {\bibfnamefont {R.}~\bibnamefont {Aditi~Simha}},\ }\bibfield  {title}
  {\enquote {\bibinfo {title} {Hydrodynamics of soft active matter},}\ }\href
  {\doibase https://doi.org/10.1103/RevModPhys.85.1143} {\bibfield  {journal}
  {\bibinfo  {journal} {Rev. Mod. Phys.}\ }\textbf {\bibinfo {volume} {85}},\
  \bibinfo {pages} {1143} (\bibinfo {year} {2013})}\BibitemShut {NoStop}%
\bibitem [{\citenamefont {Cavagna}\ \emph {et~al.}(2010)\citenamefont
  {Cavagna}, \citenamefont {Cimarelli}, \citenamefont {Giardina}, \citenamefont
  {Parisi}, \citenamefont {Santagati}, \citenamefont {Stefanini},\ and\
  \citenamefont {M.}}]{parisi}%
  \BibitemOpen
  \bibfield  {author} {\bibinfo {author} {\bibfnamefont {A.}~\bibnamefont
  {Cavagna}}, \bibinfo {author} {\bibfnamefont {A.}~\bibnamefont {Cimarelli}},
  \bibinfo {author} {\bibfnamefont {I.}~\bibnamefont {Giardina}}, \bibinfo
  {author} {\bibfnamefont {G.}~\bibnamefont {Parisi}}, \bibinfo {author}
  {\bibfnamefont {R.}~\bibnamefont {Santagati}}, \bibinfo {author}
  {\bibfnamefont {F.}~\bibnamefont {Stefanini}}, \ and\ \bibinfo {author}
  {\bibfnamefont {Viale}\ \bibnamefont {M.}},\ }\bibfield  {title} {\enquote
  {\bibinfo {title} {Scale-free correlations in starling flocks},}\ }\href
  {\doibase https://doi.org/10.1073/pnas.10057661} {\bibfield  {journal}
  {\bibinfo  {journal} {Proc. Natl. Acad. Sci., USA}\ }\textbf {\bibinfo
  {volume} {107}},\ \bibinfo {pages} {11865--11870} (\bibinfo {year}
  {2010})}\BibitemShut {NoStop}%
\bibitem [{\citenamefont {Okubo}(1986)}]{okubo}%
  \BibitemOpen
  \bibfield  {author} {\bibinfo {author} {\bibfnamefont {A.}~\bibnamefont
  {Okubo}},\ }\bibfield  {title} {\enquote {\bibinfo {title} {Dynamical aspects
  of animal grouping: Swarms, schools, flocks, and herds},}\ }\href {\doibase
  https://doi.org/10.1016/0065-227X(86)90003-1} {\bibfield  {journal} {\bibinfo
   {journal} {Adv. Biophys.}\ }\textbf {\bibinfo {volume} {22}},\ \bibinfo
  {pages} {1--94} (\bibinfo {year} {1986})}\BibitemShut {NoStop}%
\bibitem [{\citenamefont {Dombrowski}\ \emph {et~al.}(2004)\citenamefont
  {Dombrowski}, \citenamefont {Cisneros}, \citenamefont {Chatkaew},
  \citenamefont {Goldstein},\ and\ \citenamefont {Kessler}}]{dombrowski}%
  \BibitemOpen
  \bibfield  {author} {\bibinfo {author} {\bibfnamefont {C.}~\bibnamefont
  {Dombrowski}}, \bibinfo {author} {\bibfnamefont {L.}~\bibnamefont
  {Cisneros}}, \bibinfo {author} {\bibfnamefont {S.}~\bibnamefont {Chatkaew}},
  \bibinfo {author} {\bibfnamefont {R.~E.}\ \bibnamefont {Goldstein}}, \ and\
  \bibinfo {author} {\bibfnamefont {J.~O.}\ \bibnamefont {Kessler}},\
  }\bibfield  {title} {\enquote {\bibinfo {title} {Self-concentration and
  large-scale coherence in bacterial dynamics},}\ }\href@noop {} {\bibfield
  {journal} {\bibinfo  {journal} {Phys. Rev. Lett.}\ }\textbf {\bibinfo
  {volume} {93}},\ \bibinfo {pages} {098103} (\bibinfo {year}
  {2004})}\BibitemShut {NoStop}%
\bibitem [{\citenamefont {Prost}\ \emph {et~al.}(2015)\citenamefont {Prost},
  \citenamefont {J\"ulicher},\ and\ \citenamefont {Joanny}}]{joanny}%
  \BibitemOpen
  \bibfield  {author} {\bibinfo {author} {\bibfnamefont {J.}~\bibnamefont
  {Prost}}, \bibinfo {author} {\bibfnamefont {F.}~\bibnamefont {J\"ulicher}}, \
  and\ \bibinfo {author} {\bibfnamefont {J.~F.}\ \bibnamefont {Joanny}},\
  }\bibfield  {title} {\enquote {\bibinfo {title} {Active gel physics},}\
  }\href {\doibase https://doi.org/10.1038/nphys3224} {\bibfield  {journal}
  {\bibinfo  {journal} {Nature Physics}\ }\textbf {\bibinfo {volume} {11}},\
  \bibinfo {pages} {111--117} (\bibinfo {year} {2015})}\BibitemShut {NoStop}%
\bibitem [{\citenamefont {Bechinger}\ \emph {et~al.}(2016)\citenamefont
  {Bechinger}, \citenamefont {Di~Leonardo}, \citenamefont {L\"owen},
  \citenamefont {Reichhardt}, \citenamefont {Volpe},\ and\ \citenamefont
  {Volpe}}]{bechinger}%
  \BibitemOpen
  \bibfield  {author} {\bibinfo {author} {\bibfnamefont {C.}~\bibnamefont
  {Bechinger}}, \bibinfo {author} {\bibfnamefont {R.}~\bibnamefont
  {Di~Leonardo}}, \bibinfo {author} {\bibfnamefont {H.}~\bibnamefont
  {L\"owen}}, \bibinfo {author} {\bibfnamefont {C.}~\bibnamefont {Reichhardt}},
  \bibinfo {author} {\bibfnamefont {G.}~\bibnamefont {Volpe}}, \ and\ \bibinfo
  {author} {\bibfnamefont {G.}~\bibnamefont {Volpe}},\ }\bibfield  {title}
  {\enquote {\bibinfo {title} {Active particles in complex and crowded
  environments},}\ }\href {\doibase
  https://doi.org/10.1103/RevModPhys.88.045006} {\bibfield  {journal} {\bibinfo
   {journal} {Rev. Mod. Phys.}\ }\textbf {\bibinfo {volume} {88}},\ \bibinfo
  {pages} {045006} (\bibinfo {year} {2016})}\BibitemShut {NoStop}%
\bibitem [{\citenamefont {Surrey}\ \emph {et~al.}(2001)\citenamefont {Surrey},
  \citenamefont {Leibler},\ and\ \citenamefont {Karsenti}}]{surrey}%
  \BibitemOpen
  \bibfield  {author} {\bibinfo {author} {\bibfnamefont {F.~J.}\ \bibnamefont
  {Surrey}, \bibfnamefont {T.~amd~N\'ed\'elec}}, \bibinfo {author}
  {\bibfnamefont {S.}~\bibnamefont {Leibler}}, \ and\ \bibinfo {author}
  {\bibfnamefont {E.}~\bibnamefont {Karsenti}},\ }\bibfield  {title} {\enquote
  {\bibinfo {title} {Physical properties determining self-organization of
  motors and microtubules},}\ }\href {\doibase 10.1126/science.1059758}
  {\bibfield  {journal} {\bibinfo  {journal} {Science}\ }\textbf {\bibinfo
  {volume} {292}},\ \bibinfo {pages} {1167} (\bibinfo {year}
  {2001})}\BibitemShut {NoStop}%
\bibitem [{\citenamefont {Uchida}\ and\ \citenamefont
  {Golestanian}(2010)}]{ramin2}%
  \BibitemOpen
  \bibfield  {author} {\bibinfo {author} {\bibfnamefont {N.}~\bibnamefont
  {Uchida}}\ and\ \bibinfo {author} {\bibfnamefont {R.}~\bibnamefont
  {Golestanian}},\ }\bibfield  {title} {\enquote {\bibinfo {title}
  {Synchronization and collective dynamics in a carpet of microfluidic
  rotors},}\ }\href {\doibase https://doi.org/10.1103/PhysRevLett.104.178103}
  {\bibfield  {journal} {\bibinfo  {journal} {Phys. Rev. Lett.}\ }\textbf
  {\bibinfo {volume} {104}},\ \bibinfo {pages} {178103} (\bibinfo {year}
  {2010})}\BibitemShut {NoStop}%
\bibitem [{\citenamefont {Wensink}\ \emph {et~al.}(2012)\citenamefont
  {Wensink}, \citenamefont {Dunkel}, \citenamefont {Heidenreich}, \citenamefont
  {Drescher}, \citenamefont {Lowen}, \citenamefont {Goldstein},\ and\
  \citenamefont {Yeomans}}]{wensink}%
  \BibitemOpen
  \bibfield  {author} {\bibinfo {author} {\bibfnamefont {H.~H.}\ \bibnamefont
  {Wensink}}, \bibinfo {author} {\bibfnamefont {J.}~\bibnamefont {Dunkel}},
  \bibinfo {author} {\bibfnamefont {S.}~\bibnamefont {Heidenreich}}, \bibinfo
  {author} {\bibfnamefont {K.}~\bibnamefont {Drescher}}, \bibinfo {author}
  {\bibfnamefont {H.}~\bibnamefont {Lowen}}, \bibinfo {author} {\bibfnamefont
  {R.~E.}\ \bibnamefont {Goldstein}}, \ and\ \bibinfo {author} {\bibfnamefont
  {J.~M.}\ \bibnamefont {Yeomans}},\ }\bibfield  {title} {\enquote {\bibinfo
  {title} {Meso-scale turbulence in living fluids},}\ }\href {\doibase
  https://doi.org/10.1073/pnas.1202032109} {\bibfield  {journal} {\bibinfo
  {journal} {Proc. Natl. Acad. Sci. USA}\ }\textbf {\bibinfo {volume} {109}},\
  \bibinfo {pages} {14308--14313} (\bibinfo {year} {2012})}\BibitemShut
  {NoStop}%
\bibitem [{\citenamefont {Dunkel}\ \emph {et~al.}(2013)\citenamefont {Dunkel},
  \citenamefont {Heidenreich}, \citenamefont {Drescher}, \citenamefont
  {Wensink}, \citenamefont {B\"ar},\ and\ \citenamefont {Goldstein}}]{dunkel}%
  \BibitemOpen
  \bibfield  {author} {\bibinfo {author} {\bibfnamefont {J.}~\bibnamefont
  {Dunkel}}, \bibinfo {author} {\bibfnamefont {S.}~\bibnamefont {Heidenreich}},
  \bibinfo {author} {\bibfnamefont {K.}~\bibnamefont {Drescher}}, \bibinfo
  {author} {\bibfnamefont {H.~H.}\ \bibnamefont {Wensink}}, \bibinfo {author}
  {\bibfnamefont {M}~\bibnamefont {B\"ar}}, \ and\ \bibinfo {author}
  {\bibfnamefont {R.~E.}\ \bibnamefont {Goldstein}},\ }\bibfield  {title}
  {\enquote {\bibinfo {title} {Fluid dynamics of bacterial turbulence},}\
  }\href@noop {} {\bibfield  {journal} {\bibinfo  {journal} {Phys. Rev. Lett.}\
  }\textbf {\bibinfo {volume} {110}},\ \bibinfo {pages} {228102} (\bibinfo
  {year} {2013})}\BibitemShut {NoStop}%
\bibitem [{\citenamefont {Golestanian}(2009)}]{ramin}%
  \BibitemOpen
  \bibfield  {author} {\bibinfo {author} {\bibfnamefont {R.}~\bibnamefont
  {Golestanian}},\ }\bibfield  {title} {\enquote {\bibinfo {title} {Anomalous
  diffusion of symmetric and asymmetric active colloids},}\ }\href {\doibase
  https://doi.org/10.1103/PhysRevLett.102.188305} {\bibfield  {journal}
  {\bibinfo  {journal} {Phys. Rev. Lett.}\ }\textbf {\bibinfo {volume} {102}},\
  \bibinfo {pages} {188305} (\bibinfo {year} {2009})}\BibitemShut {NoStop}%
\bibitem [{\citenamefont {L\'opez}\ \emph {et~al.}(2015)\citenamefont
  {L\'opez}, \citenamefont {Gachelin}, \citenamefont {Douarche}, \citenamefont
  {Auradou},\ and\ \citenamefont {Cl\'ement}}]{clement}%
  \BibitemOpen
  \bibfield  {author} {\bibinfo {author} {\bibfnamefont {H.~M.}\ \bibnamefont
  {L\'opez}}, \bibinfo {author} {\bibfnamefont {J.}~\bibnamefont {Gachelin}},
  \bibinfo {author} {\bibfnamefont {C.}~\bibnamefont {Douarche}}, \bibinfo
  {author} {\bibfnamefont {H.}~\bibnamefont {Auradou}}, \ and\ \bibinfo
  {author} {\bibfnamefont {E.}~\bibnamefont {Cl\'ement}},\ }\bibfield  {title}
  {\enquote {\bibinfo {title} {Turning bacteria suspensions into
  superfluids},}\ }\href {\doibase
  https://doi.org/10.1103/PhysRevLett.115.028301} {\bibfield  {journal}
  {\bibinfo  {journal} {Phys. Rev. Lett.}\ }\textbf {\bibinfo {volume} {115}},\
  \bibinfo {pages} {028301} (\bibinfo {year} {2015})}\BibitemShut {NoStop}%
\bibitem [{\citenamefont {Saintillan}(2018)}]{saintillan}%
  \BibitemOpen
  \bibfield  {author} {\bibinfo {author} {\bibfnamefont {D.}~\bibnamefont
  {Saintillan}},\ }\bibfield  {title} {\enquote {\bibinfo {title} {Rheology of
  active fluids},}\ }\href {\doibase
  https://doi.org/10.1146/annurev-fluid-010816-060049} {\bibfield  {journal}
  {\bibinfo  {journal} {Annu. Rev. Fluid Mech.}\ }\textbf {\bibinfo {volume}
  {50}},\ \bibinfo {pages} {563--592} (\bibinfo {year} {2018})}\BibitemShut
  {NoStop}%
\bibitem [{\citenamefont {Cates}\ and\ \citenamefont {Tailleur}(2015)}]{mips}%
  \BibitemOpen
  \bibfield  {author} {\bibinfo {author} {\bibfnamefont {M.~E.}\ \bibnamefont
  {Cates}}\ and\ \bibinfo {author} {\bibfnamefont {J.}~\bibnamefont
  {Tailleur}},\ }\bibfield  {title} {\enquote {\bibinfo {title}
  {Motility-induced phase separation},}\ }\href {\doibase
  https://doi.org/10.1146/annurev-conmatphys-031214-014710} {\bibfield
  {journal} {\bibinfo  {journal} {Annu. Rev. Cond. Matt. Phys.}\ }\textbf
  {\bibinfo {volume} {6}},\ \bibinfo {pages} {219--244} (\bibinfo {year}
  {2015})}\BibitemShut {NoStop}%
\bibitem [{\citenamefont {Gonnella}\ \emph {et~al.}(2015)\citenamefont
  {Gonnella}, \citenamefont {Marenduzzo}, \citenamefont {Suma},\ and\
  \citenamefont {A.}}]{gonnella}%
  \BibitemOpen
  \bibfield  {author} {\bibinfo {author} {\bibfnamefont {G.}~\bibnamefont
  {Gonnella}}, \bibinfo {author} {\bibfnamefont {D.}~\bibnamefont
  {Marenduzzo}}, \bibinfo {author} {\bibfnamefont {A.}~\bibnamefont {Suma}}, \
  and\ \bibinfo {author} {\bibfnamefont {Tiribocchi}\ \bibnamefont {A.}},\
  }\bibfield  {title} {\enquote {\bibinfo {title} {Motility-induced phase
  separation and coarsening in active matter},}\ }\href {\doibase
  https://doi.org/10.1016/j.crhy.2015.05.001} {\bibfield  {journal} {\bibinfo
  {journal} {Comptes Rendus Physique}\ }\textbf {\bibinfo {volume} {16}},\
  \bibinfo {pages} {316--331} (\bibinfo {year} {2015})}\BibitemShut {NoStop}%
\bibitem [{\citenamefont {Sanchez}\ \emph {et~al.}(2012)\citenamefont
  {Sanchez}, \citenamefont {Chen}, \citenamefont {DeCamp}, \citenamefont
  {Heymann},\ and\ \citenamefont {Dogic}}]{dogic}%
  \BibitemOpen
  \bibfield  {author} {\bibinfo {author} {\bibfnamefont {T.}~\bibnamefont
  {Sanchez}}, \bibinfo {author} {\bibfnamefont {D.~T.~N.}\ \bibnamefont
  {Chen}}, \bibinfo {author} {\bibfnamefont {S.~J.}\ \bibnamefont {DeCamp}},
  \bibinfo {author} {\bibfnamefont {M.}~\bibnamefont {Heymann}}, \ and\
  \bibinfo {author} {\bibfnamefont {Z.}~\bibnamefont {Dogic}},\ }\bibfield
  {title} {\enquote {\bibinfo {title} {Spontaneous motion in hierarchically
  assembled active matter},}\ }\href {\doibase
  https://doi.org/10.1038/nature11591} {\bibfield  {journal} {\bibinfo
  {journal} {Nature}\ }\textbf {\bibinfo {volume} {491}},\ \bibinfo {pages}
  {431--434} (\bibinfo {year} {2012})}\BibitemShut {NoStop}%
\bibitem [{\citenamefont {Guillamat}\ \emph {et~al.}(2018)\citenamefont
  {Guillamat}, \citenamefont {Kos}, \citenamefont {Hardo\"uin}, \citenamefont
  {Ign\'es-Mullol}, \citenamefont {Ravnik},\ and\ \citenamefont
  {Sagu\'es}}]{sagues2}%
  \BibitemOpen
  \bibfield  {author} {\bibinfo {author} {\bibfnamefont {P.}~\bibnamefont
  {Guillamat}}, \bibinfo {author} {\bibfnamefont {Z.}~\bibnamefont {Kos}},
  \bibinfo {author} {\bibfnamefont {J.}~\bibnamefont {Hardo\"uin}}, \bibinfo
  {author} {\bibfnamefont {J.}~\bibnamefont {Ign\'es-Mullol}}, \bibinfo
  {author} {\bibfnamefont {M.}~\bibnamefont {Ravnik}}, \ and\ \bibinfo {author}
  {\bibfnamefont {F.}~\bibnamefont {Sagu\'es}},\ }\bibfield  {title} {\enquote
  {\bibinfo {title} {Active nematic emulsions},}\ }\href {\doibase
  10.1126/sciadv.aao1470} {\bibfield  {journal} {\bibinfo  {journal} {Science
  Advances}\ }\textbf {\bibinfo {volume} {4}},\ \bibinfo {pages} {eaao1470}
  (\bibinfo {year} {2018})}\BibitemShut {NoStop}%
\bibitem [{\citenamefont {Ign\'es-Mullol}\ and\ \citenamefont
  {Sagu\'es}(2020)}]{sagues3}%
  \BibitemOpen
  \bibfield  {author} {\bibinfo {author} {\bibfnamefont {J.}~\bibnamefont
  {Ign\'es-Mullol}}\ and\ \bibinfo {author} {\bibfnamefont {F.}~\bibnamefont
  {Sagu\'es}},\ }\bibfield  {title} {\enquote {\bibinfo {title} {Active,
  self-motile, and driven emulsions},}\ }\href {\doibase
  https://doi.org/10.1016/j.cocis.2020.04.007} {\bibfield  {journal} {\bibinfo
  {journal} {Curr. Op. Coll. \& Interf. Sci.}\ }\textbf {\bibinfo {volume}
  {49}},\ \bibinfo {pages} {16--26} (\bibinfo {year} {2020})}\BibitemShut
  {NoStop}%
\bibitem [{\citenamefont {Saha}\ \emph {et~al.}(2022)\citenamefont {Saha},
  \citenamefont {Das}, \citenamefont {Patra}, \citenamefont {Anilkumar},
  \citenamefont {Sil}, \citenamefont {Mayor},\ and\ \citenamefont {Rao}}]{rao}%
  \BibitemOpen
  \bibfield  {author} {\bibinfo {author} {\bibfnamefont {S.}~\bibnamefont
  {Saha}}, \bibinfo {author} {\bibfnamefont {A.}~\bibnamefont {Das}}, \bibinfo
  {author} {\bibfnamefont {C.}~\bibnamefont {Patra}}, \bibinfo {author}
  {\bibfnamefont {A.~A.}\ \bibnamefont {Anilkumar}}, \bibinfo {author}
  {\bibfnamefont {P.}~\bibnamefont {Sil}}, \bibinfo {author} {\bibfnamefont
  {S.}~\bibnamefont {Mayor}}, \ and\ \bibinfo {author} {\bibfnamefont
  {M.}~\bibnamefont {Rao}},\ }\bibfield  {title} {\enquote {\bibinfo {title}
  {Active emulsions in living cell membranes driven by contractile stresses and
  transbilayer coupling},}\ }\href {\doibase
  https://doi.org/10.1073/pnas.2123056119} {\bibfield  {journal} {\bibinfo
  {journal} {Proc. Natl. Acad. Sci. USA}\ }\textbf {\bibinfo {volume} {119}},\
  \bibinfo {pages} {e2123056119} (\bibinfo {year} {2022})}\BibitemShut
  {NoStop}%
\bibitem [{\citenamefont {Sakamoto}\ \emph {et~al.}(2022)\citenamefont
  {Sakamoto}, \citenamefont {Izri}, \citenamefont {Shimamoto},\ and\
  \citenamefont {Maeda}}]{maeda}%
  \BibitemOpen
  \bibfield  {author} {\bibinfo {author} {\bibfnamefont {R.}~\bibnamefont
  {Sakamoto}}, \bibinfo {author} {\bibfnamefont {Z.}~\bibnamefont {Izri}},
  \bibinfo {author} {\bibfnamefont {Y.}~\bibnamefont {Shimamoto}}, \ and\
  \bibinfo {author} {\bibfnamefont {Y.~T.}\ \bibnamefont {Maeda}},\ }\bibfield
  {title} {\enquote {\bibinfo {title} {Geometric trade-off between contractile
  force and viscous drag determines the actomyosin-based motility of a
  cell-sized droplet},}\ }\href {\doibase
  https://doi.org/10.1073/pnas.2121147119} {\bibfield  {journal} {\bibinfo
  {journal} {Proc. Natl. Acad. Sci. USA}\ }\textbf {\bibinfo {volume} {119}},\
  \bibinfo {pages} {e2121147119} (\bibinfo {year} {2022})}\BibitemShut
  {NoStop}%
\bibitem [{\citenamefont {K\"ohler}\ \emph {et~al.}(2011)\citenamefont
  {K\"ohler}, \citenamefont {Schaller},\ and\ \citenamefont {Bausch}}]{koler}%
  \BibitemOpen
  \bibfield  {author} {\bibinfo {author} {\bibfnamefont {S.}~\bibnamefont
  {K\"ohler}}, \bibinfo {author} {\bibfnamefont {V.}~\bibnamefont {Schaller}},
  \ and\ \bibinfo {author} {\bibfnamefont {A.~R.}\ \bibnamefont {Bausch}},\
  }\bibfield  {title} {\enquote {\bibinfo {title} {Structure formation in
  active networks},}\ }\href@noop {} {\bibfield  {journal} {\bibinfo  {journal}
  {Nature Materials}\ }\textbf {\bibinfo {volume} {10}},\ \bibinfo {pages}
  {462--468} (\bibinfo {year} {2011})}\BibitemShut {NoStop}%
\bibitem [{\citenamefont {Loisel}\ \emph {et~al.}(1999)\citenamefont {Loisel},
  \citenamefont {Boujemaa}, \citenamefont {Pantaloni},\ and\ \citenamefont
  {Carlier}}]{loisel}%
  \BibitemOpen
  \bibfield  {author} {\bibinfo {author} {\bibfnamefont {T.}~\bibnamefont
  {Loisel}}, \bibinfo {author} {\bibfnamefont {R.}~\bibnamefont {Boujemaa}},
  \bibinfo {author} {\bibfnamefont {D.}~\bibnamefont {Pantaloni}}, \ and\
  \bibinfo {author} {\bibfnamefont {M.~F.}\ \bibnamefont {Carlier}},\
  }\bibfield  {title} {\enquote {\bibinfo {title} {Reconstitution of
  actin-based movement using pure proteins},}\ }\href@noop {} {\bibfield
  {journal} {\bibinfo  {journal} {Nature}\ }\textbf {\bibinfo {volume} {401}},\
  \bibinfo {pages} {603--616} (\bibinfo {year} {1999})}\BibitemShut {NoStop}%
\bibitem [{\citenamefont {Sumino}\ \emph {et~al.}(2012)\citenamefont {Sumino},
  \citenamefont {Nagai}, \citenamefont {Shitaka}, \citenamefont {Yoshikawa},
  \citenamefont {Chat\'e},\ and\ \citenamefont {Oiwa}}]{sumino}%
  \BibitemOpen
  \bibfield  {author} {\bibinfo {author} {\bibfnamefont {Y.}~\bibnamefont
  {Sumino}}, \bibinfo {author} {\bibfnamefont {K.~H.}\ \bibnamefont {Nagai}},
  \bibinfo {author} {\bibfnamefont {Y.}~\bibnamefont {Shitaka}}, \bibinfo
  {author} {\bibfnamefont {K.}~\bibnamefont {Yoshikawa}}, \bibinfo {author}
  {\bibfnamefont {H.}~\bibnamefont {Chat\'e}}, \ and\ \bibinfo {author}
  {\bibfnamefont {K.}~\bibnamefont {Oiwa}},\ }\bibfield  {title} {\enquote
  {\bibinfo {title} {Large-scale vortex lattice emerging from collectively
  moving microtubules},}\ }\href {\doibase https://doi.org/10.1038/nature10874}
  {\bibfield  {journal} {\bibinfo  {journal} {Nature}\ }\textbf {\bibinfo
  {volume} {493}},\ \bibinfo {pages} {448--452} (\bibinfo {year}
  {2012})}\BibitemShut {NoStop}%
\bibitem [{\citenamefont {Ramaswamy}(2010)}]{sriram}%
  \BibitemOpen
  \bibfield  {author} {\bibinfo {author} {\bibfnamefont {S.}~\bibnamefont
  {Ramaswamy}},\ }\bibfield  {title} {\enquote {\bibinfo {title} {The mechanics
  and statistics of active matter},}\ }\href {\doibase
  https://doi.org/10.1146/annurev-conmatphys-070909-104101} {\bibfield
  {journal} {\bibinfo  {journal} {Annu. Rev. Cond. Mat. Phys.}\ }\textbf
  {\bibinfo {volume} {1}},\ \bibinfo {pages} {323--345} (\bibinfo {year}
  {2010})}\BibitemShut {NoStop}%
\bibitem [{\citenamefont {Simha}\ and\ \citenamefont
  {Ramaswamy}(2002)}]{ramaswamy}%
  \BibitemOpen
  \bibfield  {author} {\bibinfo {author} {\bibfnamefont {R.~A.}\ \bibnamefont
  {Simha}}\ and\ \bibinfo {author} {\bibfnamefont {S.}~\bibnamefont
  {Ramaswamy}},\ }\bibfield  {title} {\enquote {\bibinfo {title} {Hydrodynamic
  fluctuations and instabilities in ordered suspensions of self-propelled
  particles},}\ }\href {\doibase https://doi.org/10.1103/PhysRevLett.89.058101}
  {\bibfield  {journal} {\bibinfo  {journal} {Phys. Rev. Lett.}\ }\textbf
  {\bibinfo {volume} {89}},\ \bibinfo {pages} {058101} (\bibinfo {year}
  {2002})}\BibitemShut {NoStop}%
\bibitem [{\citenamefont {Kozhukhov}\ and\ \citenamefont
  {Shendruk}(2022)}]{shendruk}%
  \BibitemOpen
  \bibfield  {author} {\bibinfo {author} {\bibfnamefont {T.}~\bibnamefont
  {Kozhukhov}}\ and\ \bibinfo {author} {\bibfnamefont {T.~N.}\ \bibnamefont
  {Shendruk}},\ }\bibfield  {title} {\enquote {\bibinfo {title} {Mesoscopic
  simulations of active nematics},}\ }\href {\doibase 10.1126/sciadv.abo5788}
  {\bibfield  {journal} {\bibinfo  {journal} {Sci. Advances}\ }\textbf
  {\bibinfo {volume} {8}},\ \bibinfo {pages} {34} (\bibinfo {year}
  {2022})}\BibitemShut {NoStop}%
\bibitem [{\citenamefont {Tjhung}\ \emph {et~al.}(2012)\citenamefont {Tjhung},
  \citenamefont {Marenduzzo},\ and\ \citenamefont {Cates}}]{tjhung1}%
  \BibitemOpen
  \bibfield  {author} {\bibinfo {author} {\bibfnamefont {E.}~\bibnamefont
  {Tjhung}}, \bibinfo {author} {\bibfnamefont {D.}~\bibnamefont {Marenduzzo}},
  \ and\ \bibinfo {author} {\bibfnamefont {M.~E.}\ \bibnamefont {Cates}},\
  }\bibfield  {title} {\enquote {\bibinfo {title} {Spontaneous symmetry
  breaking in active droplets provides a generic route to motility},}\ }\href
  {\doibase https://doi.org/10.1073/pnas.1200843109} {\bibfield  {journal}
  {\bibinfo  {journal} {Proc. Nat. Acad. Sci., USA}\ }\textbf {\bibinfo
  {volume} {109}},\ \bibinfo {pages} {12381--12386} (\bibinfo {year}
  {2012})}\BibitemShut {NoStop}%
\bibitem [{\citenamefont {Giomi}\ and\ \citenamefont {DeSimone}(2014)}]{giomi}%
  \BibitemOpen
  \bibfield  {author} {\bibinfo {author} {\bibfnamefont {L.}~\bibnamefont
  {Giomi}}\ and\ \bibinfo {author} {\bibfnamefont {A.}~\bibnamefont
  {DeSimone}},\ }\bibfield  {title} {\enquote {\bibinfo {title} {Spontaneous
  division and motility in active nematic droplets},}\ }\href {\doibase
  https://doi.org/10.1103/PhysRevLett.112.147802} {\bibfield  {journal}
  {\bibinfo  {journal} {Phys. Rev. Lett.}\ }\textbf {\bibinfo {volume} {112}},\
  \bibinfo {pages} {147802} (\bibinfo {year} {2014})}\BibitemShut {NoStop}%
\bibitem [{\citenamefont {Khoromskaia}\ and\ \citenamefont
  {Alexander}(2015)}]{gareth_pre}%
  \BibitemOpen
  \bibfield  {author} {\bibinfo {author} {\bibfnamefont {D.}~\bibnamefont
  {Khoromskaia}}\ and\ \bibinfo {author} {\bibfnamefont {G.~P.}\ \bibnamefont
  {Alexander}},\ }\bibfield  {title} {\enquote {\bibinfo {title} {Motility of
  active fluid drops on surfaces},}\ }\href@noop {} {\bibfield  {journal}
  {\bibinfo  {journal} {Phys. Rev. E}\ }\textbf {\bibinfo {volume} {92}},\
  \bibinfo {pages} {062311} (\bibinfo {year} {2015})}\BibitemShut {NoStop}%
\bibitem [{\citenamefont {Zwicker}\ \emph {et~al.}(2017)\citenamefont
  {Zwicker}, \citenamefont {Seyboldt}, \citenamefont {Weber}, \citenamefont
  {Hyman},\ and\ \citenamefont {Julicher}}]{zwicker}%
  \BibitemOpen
  \bibfield  {author} {\bibinfo {author} {\bibfnamefont {D.}~\bibnamefont
  {Zwicker}}, \bibinfo {author} {\bibfnamefont {R.}~\bibnamefont {Seyboldt}},
  \bibinfo {author} {\bibfnamefont {C.~A.}\ \bibnamefont {Weber}}, \bibinfo
  {author} {\bibfnamefont {A.~A.}\ \bibnamefont {Hyman}}, \ and\ \bibinfo
  {author} {\bibfnamefont {F.}~\bibnamefont {Julicher}},\ }\bibfield  {title}
  {\enquote {\bibinfo {title} {Growth and division of active droplets provides
  a model for protocells},}\ }\href@noop {} {\bibfield  {journal} {\bibinfo
  {journal} {Nat. Phys.}\ }\textbf {\bibinfo {volume} {13}},\ \bibinfo {pages}
  {408--413} (\bibinfo {year} {2017})}\BibitemShut {NoStop}%
\bibitem [{\citenamefont {Saw}\ \emph {et~al.}(2017)\citenamefont {Saw},
  \citenamefont {Doostmohammadi}, \citenamefont {Nier}, \citenamefont
  {Kocgozlu}, \citenamefont {Thampi}, \citenamefont {Toyama}, \citenamefont
  {Marcq}, \citenamefont {Lim}, \citenamefont {Yeomans},\ and\ \citenamefont
  {Ladoux}}]{yeom_nat}%
  \BibitemOpen
  \bibfield  {author} {\bibinfo {author} {\bibfnamefont {T.~B.}\ \bibnamefont
  {Saw}}, \bibinfo {author} {\bibfnamefont {A.}~\bibnamefont {Doostmohammadi}},
  \bibinfo {author} {\bibfnamefont {V.}~\bibnamefont {Nier}}, \bibinfo {author}
  {\bibfnamefont {L.}~\bibnamefont {Kocgozlu}}, \bibinfo {author}
  {\bibfnamefont {S.}~\bibnamefont {Thampi}}, \bibinfo {author} {\bibfnamefont
  {Y.}~\bibnamefont {Toyama}}, \bibinfo {author} {\bibfnamefont
  {P.}~\bibnamefont {Marcq}}, \bibinfo {author} {\bibfnamefont {C.~T.}\
  \bibnamefont {Lim}}, \bibinfo {author} {\bibfnamefont {J.~M.}\ \bibnamefont
  {Yeomans}}, \ and\ \bibinfo {author} {\bibfnamefont {B.}~\bibnamefont
  {Ladoux}},\ }\bibfield  {title} {\enquote {\bibinfo {title} {Topological
  defects in epithelia govern cell death and extrusion},}\ }\href {\doibase
  https://doi.org/10.1038/nature21718} {\bibfield  {journal} {\bibinfo
  {journal} {Nature}\ }\textbf {\bibinfo {volume} {544}},\ \bibinfo {pages}
  {212--216} (\bibinfo {year} {2017})}\BibitemShut {NoStop}%
\bibitem [{\citenamefont {Tiribocchi}\ \emph {et~al.}(2023)\citenamefont
  {Tiribocchi}, \citenamefont {Durve}, \citenamefont {Lauricella},
  \citenamefont {Montessori}, \citenamefont {Marenduzzo},\ and\ \citenamefont
  {Succi}}]{tiribocchi_nat}%
  \BibitemOpen
  \bibfield  {author} {\bibinfo {author} {\bibfnamefont {A.}~\bibnamefont
  {Tiribocchi}}, \bibinfo {author} {\bibfnamefont {M.}~\bibnamefont {Durve}},
  \bibinfo {author} {\bibfnamefont {M.}~\bibnamefont {Lauricella}}, \bibinfo
  {author} {\bibfnamefont {A.}~\bibnamefont {Montessori}}, \bibinfo {author}
  {\bibfnamefont {D.}~\bibnamefont {Marenduzzo}}, \ and\ \bibinfo {author}
  {\bibfnamefont {S.}~\bibnamefont {Succi}},\ }\bibfield  {title} {\enquote
  {\bibinfo {title} {The crucial role of adhesion in the transmigration of
  active droplets through interstitial orifices},}\ }\href {\doibase
  https://doi.org/10.1038/s41467-023-36656-0} {\bibfield  {journal} {\bibinfo
  {journal} {Nature Commun.}\ }\textbf {\bibinfo {volume} {14}},\ \bibinfo
  {pages} {1096} (\bibinfo {year} {2023})}\BibitemShut {NoStop}%
\bibitem [{\citenamefont {Carenza}\ \emph {et~al.}(2019)\citenamefont
  {Carenza}, \citenamefont {Gonnella}, \citenamefont {Marenduzzo},\ and\
  \citenamefont {Negro}}]{carenza_pnas}%
  \BibitemOpen
  \bibfield  {author} {\bibinfo {author} {\bibfnamefont {L.~N.}\ \bibnamefont
  {Carenza}}, \bibinfo {author} {\bibfnamefont {G.}~\bibnamefont {Gonnella}},
  \bibinfo {author} {\bibfnamefont {D.}~\bibnamefont {Marenduzzo}}, \ and\
  \bibinfo {author} {\bibfnamefont {G.}~\bibnamefont {Negro}},\ }\bibfield
  {title} {\enquote {\bibinfo {title} {Rotation and propulsion in 3d active
  chiral droplets},}\ }\href@noop {} {\bibfield  {journal} {\bibinfo  {journal}
  {Proc. Natl. Acad. Sci., USA}\ }\textbf {\bibinfo {volume} {116}},\ \bibinfo
  {pages} {22065--22070} (\bibinfo {year} {2019})}\BibitemShut {NoStop}%
\bibitem [{\citenamefont {Loisy}\ \emph {et~al.}(2019)\citenamefont {Loisy},
  \citenamefont {Eggers},\ and\ \citenamefont {Liverpool}}]{liverpool_prl}%
  \BibitemOpen
  \bibfield  {author} {\bibinfo {author} {\bibfnamefont {A.}~\bibnamefont
  {Loisy}}, \bibinfo {author} {\bibfnamefont {J.}~\bibnamefont {Eggers}}, \
  and\ \bibinfo {author} {\bibfnamefont {T.~B.}\ \bibnamefont {Liverpool}},\
  }\bibfield  {title} {\enquote {\bibinfo {title} {Tractionless self-propulsion
  of active drops},}\ }\href {\doibase
  DOI:https://doi.org/10.1103/PhysRevLett.123.248006} {\bibfield  {journal}
  {\bibinfo  {journal} {Phys. Rev. Lett.}\ }\textbf {\bibinfo {volume} {123}},\
  \bibinfo {pages} {248006} (\bibinfo {year} {2019})}\BibitemShut {NoStop}%
\bibitem [{\citenamefont {Hokmabad}\ \emph {et~al.}(2019)\citenamefont
  {Hokmabad}, \citenamefont {Baldwin}, \citenamefont {Kr\"uger}, \citenamefont
  {Bahr},\ and\ \citenamefont {Maass}}]{maass}%
  \BibitemOpen
  \bibfield  {author} {\bibinfo {author} {\bibfnamefont {B.~V.}\ \bibnamefont
  {Hokmabad}}, \bibinfo {author} {\bibfnamefont {K.~A.}\ \bibnamefont
  {Baldwin}}, \bibinfo {author} {\bibfnamefont {C.}~\bibnamefont {Kr\"uger}},
  \bibinfo {author} {\bibfnamefont {C.}~\bibnamefont {Bahr}}, \ and\ \bibinfo
  {author} {\bibfnamefont {C.~C.}\ \bibnamefont {Maass}},\ }\bibfield  {title}
  {\enquote {\bibinfo {title} {Topological stabilization and dynamics of
  self-propelling nematic shells},}\ }\href {\doibase
  https://doi.org/10.1103/PhysRevLett.123.178003} {\bibfield  {journal}
  {\bibinfo  {journal} {Phys. Rev. Lett.}\ }\textbf {\bibinfo {volume} {123}},\
  \bibinfo {pages} {178003} (\bibinfo {year} {2019})}\BibitemShut {NoStop}%
\bibitem [{\citenamefont {Maass}\ \emph {et~al.}(2015)\citenamefont {Maass},
  \citenamefont {Kr\"uger}, \citenamefont {Herminghaus},\ and\ \citenamefont
  {Bahr}}]{maass2}%
  \BibitemOpen
  \bibfield  {author} {\bibinfo {author} {\bibfnamefont {C.~C.}\ \bibnamefont
  {Maass}}, \bibinfo {author} {\bibfnamefont {C.}~\bibnamefont {Kr\"uger}},
  \bibinfo {author} {\bibfnamefont {S.}~\bibnamefont {Herminghaus}}, \ and\
  \bibinfo {author} {\bibfnamefont {C.}~\bibnamefont {Bahr}},\ }\bibfield
  {title} {\enquote {\bibinfo {title} {Swimming droplets},}\ }\href@noop {}
  {\bibfield  {journal} {\bibinfo  {journal} {Annu. Rev. Cond. Mat. Phys.}\
  }\textbf {\bibinfo {volume} {7}},\ \bibinfo {pages} {171--193} (\bibinfo
  {year} {2015})}\BibitemShut {NoStop}%
\bibitem [{\citenamefont {Ziebert}\ \emph {et~al.}(2012)\citenamefont
  {Ziebert}, \citenamefont {Swaminathan},\ and\ \citenamefont
  {Aranson}}]{aranson1}%
  \BibitemOpen
  \bibfield  {author} {\bibinfo {author} {\bibfnamefont {F.}~\bibnamefont
  {Ziebert}}, \bibinfo {author} {\bibfnamefont {S.}~\bibnamefont
  {Swaminathan}}, \ and\ \bibinfo {author} {\bibfnamefont {I.~S.}\ \bibnamefont
  {Aranson}},\ }\bibfield  {title} {\enquote {\bibinfo {title} {Model for
  self-polarization and motility of keratocyte fragments},}\ }\href@noop {}
  {\bibfield  {journal} {\bibinfo  {journal} {J. R. Soc. Interface}\ }\textbf
  {\bibinfo {volume} {9}},\ \bibinfo {pages} {1084--1092} (\bibinfo {year}
  {2012})}\BibitemShut {NoStop}%
\bibitem [{\citenamefont {L\"ober}\ \emph {et~al.}(2015)\citenamefont
  {L\"ober}, \citenamefont {Ziebert},\ and\ \citenamefont
  {Aranson}}]{aranson2}%
  \BibitemOpen
  \bibfield  {author} {\bibinfo {author} {\bibfnamefont {J.}~\bibnamefont
  {L\"ober}}, \bibinfo {author} {\bibfnamefont {F.}~\bibnamefont {Ziebert}}, \
  and\ \bibinfo {author} {\bibfnamefont {I.~S.}\ \bibnamefont {Aranson}},\
  }\bibfield  {title} {\enquote {\bibinfo {title} {Collisions of deformable
  cells lead to collective migration},}\ }\href@noop {} {\bibfield  {journal}
  {\bibinfo  {journal} {Sci. Rep.}\ }\textbf {\bibinfo {volume} {5}},\ \bibinfo
  {pages} {1--7} (\bibinfo {year} {2015})}\BibitemShut {NoStop}%
\bibitem [{\citenamefont {Ziebert}\ and\ \citenamefont
  {Aranson}(2016)}]{aranson3}%
  \BibitemOpen
  \bibfield  {author} {\bibinfo {author} {\bibfnamefont {F.}~\bibnamefont
  {Ziebert}}\ and\ \bibinfo {author} {\bibfnamefont {I.~S.}\ \bibnamefont
  {Aranson}},\ }\bibfield  {title} {\enquote {\bibinfo {title} {Computational
  approaches to substrate-based cell motility},}\ }\href {\doibase
  https://doi.org/10.1038/npjcompumats.2016.19} {\bibfield  {journal} {\bibinfo
   {journal} {npj Comput. Mater.}\ }\textbf {\bibinfo {volume} {2}},\ \bibinfo
  {pages} {16019} (\bibinfo {year} {2016})}\BibitemShut {NoStop}%
\bibitem [{\citenamefont {Ruske}\ and\ \citenamefont {Yeomans}(2021)}]{ruske}%
  \BibitemOpen
  \bibfield  {author} {\bibinfo {author} {\bibfnamefont {L.~J.}\ \bibnamefont
  {Ruske}}\ and\ \bibinfo {author} {\bibfnamefont {J.~M.}\ \bibnamefont
  {Yeomans}},\ }\bibfield  {title} {\enquote {\bibinfo {title} {Morphology of
  active deformable 3d droplets},}\ }\href {\doibase
  DOI:https://doi.org/10.1103/PhysRevX.11.021001} {\bibfield  {journal}
  {\bibinfo  {journal} {Phys. Rev. X}\ }\textbf {\bibinfo {volume} {11}},\
  \bibinfo {pages} {021001} (\bibinfo {year} {2021})}\BibitemShut {NoStop}%
\bibitem [{\citenamefont {De~Magistris}\ \emph {et~al.}(2014)\citenamefont
  {De~Magistris}, \citenamefont {Tiribocchi}, \citenamefont {Whithfield},
  \citenamefont {Hawkins}, \citenamefont {Cates},\ and\ \citenamefont
  {Marenduzzo}}]{demagistris}%
  \BibitemOpen
  \bibfield  {author} {\bibinfo {author} {\bibfnamefont {G.}~\bibnamefont
  {De~Magistris}}, \bibinfo {author} {\bibfnamefont {A.}~\bibnamefont
  {Tiribocchi}}, \bibinfo {author} {\bibfnamefont {C.~A.}\ \bibnamefont
  {Whithfield}}, \bibinfo {author} {\bibfnamefont {R.~J.}\ \bibnamefont
  {Hawkins}}, \bibinfo {author} {\bibfnamefont {M.~E.}\ \bibnamefont {Cates}},
  \ and\ \bibinfo {author} {\bibfnamefont {D.}~\bibnamefont {Marenduzzo}},\
  }\bibfield  {title} {\enquote {\bibinfo {title} {Spontaneous motility of
  passive emulsion droplets in polar active gels},}\ }\href {\doibase
  https://doi.org/10.1039/C4SM00937A} {\bibfield  {journal} {\bibinfo
  {journal} {Soft Matter}\ }\textbf {\bibinfo {volume} {10}},\ \bibinfo {pages}
  {7826--7837} (\bibinfo {year} {2014})}\BibitemShut {NoStop}%
\bibitem [{\citenamefont {Poulin}\ \emph {et~al.}(97)\citenamefont {Poulin},
  \citenamefont {Stark}, \citenamefont {Lubensky},\ and\ \citenamefont
  {Weitz}}]{poulin}%
  \BibitemOpen
  \bibfield  {author} {\bibinfo {author} {\bibfnamefont {P.}~\bibnamefont
  {Poulin}}, \bibinfo {author} {\bibfnamefont {H.}~\bibnamefont {Stark}},
  \bibinfo {author} {\bibfnamefont {T.~C.}\ \bibnamefont {Lubensky}}, \ and\
  \bibinfo {author} {\bibfnamefont {D.~A.}\ \bibnamefont {Weitz}},\ }\bibfield
  {title} {\enquote {\bibinfo {title} {Novel colloidal interactions in
  anisotropic fluids},}\ }\href@noop {} {\bibfield  {journal} {\bibinfo
  {journal} {Science}\ }\textbf {\bibinfo {volume} {275}},\ \bibinfo {pages}
  {1770--1773} (\bibinfo {year} {97})}\BibitemShut {NoStop}%
\bibitem [{\citenamefont {Angelani}\ and\ \citenamefont
  {Di~Leonardo}(2010)}]{dileonardo}%
  \BibitemOpen
  \bibfield  {author} {\bibinfo {author} {\bibfnamefont {L.}~\bibnamefont
  {Angelani}}\ and\ \bibinfo {author} {\bibfnamefont {R.}~\bibnamefont
  {Di~Leonardo}},\ }\bibfield  {title} {\enquote {\bibinfo {title}
  {Geometrically biased random walks in bacteria-driven micro-shuttles},}\
  }\href {\doibase 10.1088/1367-2630/12/11/113017} {\bibfield  {journal}
  {\bibinfo  {journal} {New Journ. Phys.}\ }\textbf {\bibinfo {volume} {12}},\
  \bibinfo {pages} {113017} (\bibinfo {year} {2010})}\BibitemShut {NoStop}%
\bibitem [{\citenamefont {Di~Leonardo}\ \emph {et~al.}(2010)\citenamefont
  {Di~Leonardo}, \citenamefont {Angelani}, \citenamefont {Dell'aArciprete},
  \citenamefont {Ruocco}, \citenamefont {Iebba}, \citenamefont {Schippa},
  \citenamefont {Conte}, \citenamefont {Mecarini}, \citenamefont {De~Angelis},\
  and\ \citenamefont {Di~Fabrizio}}]{dileonardo2}%
  \BibitemOpen
  \bibfield  {author} {\bibinfo {author} {\bibfnamefont {R.}~\bibnamefont
  {Di~Leonardo}}, \bibinfo {author} {\bibfnamefont {L.}~\bibnamefont
  {Angelani}}, \bibinfo {author} {\bibfnamefont {D.}~\bibnamefont
  {Dell'aArciprete}}, \bibinfo {author} {\bibfnamefont {G.}~\bibnamefont
  {Ruocco}}, \bibinfo {author} {\bibfnamefont {V.}~\bibnamefont {Iebba}},
  \bibinfo {author} {\bibfnamefont {S.}~\bibnamefont {Schippa}}, \bibinfo
  {author} {\bibfnamefont {M.~P.}\ \bibnamefont {Conte}}, \bibinfo {author}
  {\bibfnamefont {F.}~\bibnamefont {Mecarini}}, \bibinfo {author}
  {\bibfnamefont {F.}~\bibnamefont {De~Angelis}}, \ and\ \bibinfo {author}
  {\bibfnamefont {F.}~\bibnamefont {Di~Fabrizio}},\ }\bibfield  {title}
  {\enquote {\bibinfo {title} {Bacterial ratchet motors},}\ }\href {\doibase
  https://doi.org/10.1073/pnas.0910426107} {\bibfield  {journal} {\bibinfo
  {journal} {Proc. Natl. Acad. Sci., USA}\ }\textbf {\bibinfo {volume} {107}},\
  \bibinfo {pages} {9541--9545} (\bibinfo {year} {2010})}\BibitemShut {NoStop}%
\bibitem [{\citenamefont {Sokolov}\ \emph {et~al.}(2009)\citenamefont
  {Sokolov}, \citenamefont {Apodaca}, \citenamefont {Grzybowski},\ and\
  \citenamefont {Aranson}}]{sokolov}%
  \BibitemOpen
  \bibfield  {author} {\bibinfo {author} {\bibfnamefont {A.}~\bibnamefont
  {Sokolov}}, \bibinfo {author} {\bibfnamefont {M.~M.}\ \bibnamefont
  {Apodaca}}, \bibinfo {author} {\bibfnamefont {B.~A.}\ \bibnamefont
  {Grzybowski}}, \ and\ \bibinfo {author} {\bibfnamefont {I.~D.}\ \bibnamefont
  {Aranson}},\ }\bibfield  {title} {\enquote {\bibinfo {title} {Swimming
  bacteria power microscopic gears},}\ }\href {\doibase
  https://doi.org/10.1073/pnas.0913015107} {\bibfield  {journal} {\bibinfo
  {journal} {Proc. Natl. Acad. Sci., USA}\ }\textbf {\bibinfo {volume} {107}},\
  \bibinfo {pages} {969--974} (\bibinfo {year} {2009})}\BibitemShut {NoStop}%
\bibitem [{\citenamefont {Nikola}\ \emph {et~al.}(2016)\citenamefont {Nikola},
  \citenamefont {Solon}, \citenamefont {Kafri}, \citenamefont {Tailleur},\ and\
  \citenamefont {Voituriez}}]{nikola}%
  \BibitemOpen
  \bibfield  {author} {\bibinfo {author} {\bibfnamefont {N.}~\bibnamefont
  {Nikola}}, \bibinfo {author} {\bibfnamefont {A.~P.}\ \bibnamefont {Solon}},
  \bibinfo {author} {\bibfnamefont {M.}~\bibnamefont {Kafri}, \bibfnamefont
  {Y.~an~Kardar}}, \bibinfo {author} {\bibfnamefont {J.}~\bibnamefont
  {Tailleur}}, \ and\ \bibinfo {author} {\bibfnamefont {R.}~\bibnamefont
  {Voituriez}},\ }\bibfield  {title} {\enquote {\bibinfo {title} {Active
  particles with soft and curved walls: Equation of state, ratchets, and
  instabilities},}\ }\href {\doibase
  https://doi.org/10.1103/PhysRevLett.117.098001} {\bibfield  {journal}
  {\bibinfo  {journal} {Phys. Rev. Lett.}\ }\textbf {\bibinfo {volume} {117}},\
  \bibinfo {pages} {098001} (\bibinfo {year} {2016})}\BibitemShut {NoStop}%
\bibitem [{\citenamefont {Aporvari}\ \emph {et~al.}(2020)\citenamefont
  {Aporvari}, \citenamefont {Utkur}, \citenamefont {Saritas}, \citenamefont
  {Volpe},\ and\ \citenamefont {Stenhammar}}]{joakim}%
  \BibitemOpen
  \bibfield  {author} {\bibinfo {author} {\bibfnamefont {M.~S.}\ \bibnamefont
  {Aporvari}}, \bibinfo {author} {\bibfnamefont {M.}~\bibnamefont {Utkur}},
  \bibinfo {author} {\bibfnamefont {E.~U.}\ \bibnamefont {Saritas}}, \bibinfo
  {author} {\bibfnamefont {G.}~\bibnamefont {Volpe}}, \ and\ \bibinfo {author}
  {\bibfnamefont {J.}~\bibnamefont {Stenhammar}},\ }\bibfield  {title}
  {\enquote {\bibinfo {title} {Anisotropic dynamics of a self-assembled
  colloidal chain in an active bath},}\ }\href {\doibase
  https://doi.org/10.1039/D0SM00318B} {\bibfield  {journal} {\bibinfo
  {journal} {Soft Matter}\ }\textbf {\bibinfo {volume} {16}},\ \bibinfo {pages}
  {5609--5614} (\bibinfo {year} {2020})}\BibitemShut {NoStop}%
\bibitem [{\citenamefont {Petit}\ \emph {et~al.}(2016)\citenamefont {Petit},
  \citenamefont {Polenz}, \citenamefont {Baret}, \citenamefont {Herminghaus},\
  and\ \citenamefont {B\"aumchen}}]{petit}%
  \BibitemOpen
  \bibfield  {author} {\bibinfo {author} {\bibfnamefont {J.}~\bibnamefont
  {Petit}}, \bibinfo {author} {\bibfnamefont {I.}~\bibnamefont {Polenz}},
  \bibinfo {author} {\bibfnamefont {J.~C.}\ \bibnamefont {Baret}}, \bibinfo
  {author} {\bibfnamefont {S.}~\bibnamefont {Herminghaus}}, \ and\ \bibinfo
  {author} {\bibfnamefont {O.}~\bibnamefont {B\"aumchen}},\ }\bibfield  {title}
  {\enquote {\bibinfo {title} {Vesicles-on-a-chip: A universal microfluidic
  platform for the assembly of liposomes and polymersomes},}\ }\href@noop {}
  {\bibfield  {journal} {\bibinfo  {journal} {Eur. Phys. Journ. E}\ }\textbf
  {\bibinfo {volume} {39}},\ \bibinfo {pages} {59} (\bibinfo {year}
  {2016})}\BibitemShut {NoStop}%
\bibitem [{\citenamefont {Diotallevi}\ \emph {et~al.}(2009)\citenamefont
  {Diotallevi}, \citenamefont {Biferale}, \citenamefont {Chibbaro},
  \citenamefont {Lamura}, \citenamefont {Pontrelli}, \citenamefont
  {Sbragaglia}, \citenamefont {Succi},\ and\ \citenamefont {Toschi}}]{diota}%
  \BibitemOpen
  \bibfield  {author} {\bibinfo {author} {\bibfnamefont {F.}~\bibnamefont
  {Diotallevi}}, \bibinfo {author} {\bibfnamefont {L.}~\bibnamefont
  {Biferale}}, \bibinfo {author} {\bibfnamefont {S.}~\bibnamefont {Chibbaro}},
  \bibinfo {author} {\bibfnamefont {A.}~\bibnamefont {Lamura}}, \bibinfo
  {author} {\bibfnamefont {G.}~\bibnamefont {Pontrelli}}, \bibinfo {author}
  {\bibfnamefont {M.}~\bibnamefont {Sbragaglia}}, \bibinfo {author}
  {\bibfnamefont {S.}~\bibnamefont {Succi}}, \ and\ \bibinfo {author}
  {\bibfnamefont {F.}~\bibnamefont {Toschi}},\ }\bibfield  {title} {\enquote
  {\bibinfo {title} {Capillary filling using lattice boltzmann equations: the
  case of multi-phase flows},}\ }\href@noop {} {\bibfield  {journal} {\bibinfo
  {journal} {Eur. Phys. J. Spec. Top.}\ }\textbf {\bibinfo {volume} {166}},\
  \bibinfo {pages} {111--116} (\bibinfo {year} {2009})}\BibitemShut {NoStop}%
\bibitem [{\citenamefont {Marenduzzo}\ \emph {et~al.}(2008)\citenamefont
  {Marenduzzo}, \citenamefont {Orlandini}, \citenamefont {Cates},\ and\
  \citenamefont {Yeomans}}]{mare_yeom}%
  \BibitemOpen
  \bibfield  {author} {\bibinfo {author} {\bibfnamefont {D.}~\bibnamefont
  {Marenduzzo}}, \bibinfo {author} {\bibfnamefont {E.}~\bibnamefont
  {Orlandini}}, \bibinfo {author} {\bibfnamefont {M.~E.}\ \bibnamefont
  {Cates}}, \ and\ \bibinfo {author} {\bibfnamefont {J.~M.}\ \bibnamefont
  {Yeomans}},\ }\bibfield  {title} {\enquote {\bibinfo {title} {Lattice
  boltzmann simulations of spontaneous flow in active liquid crystals: The role
  of boundary conditions},}\ }\href@noop {} {\bibfield  {journal} {\bibinfo
  {journal} {Jour. Non-Newt. Fluid Mech.}\ }\textbf {\bibinfo {volume} {149}},\
  \bibinfo {pages} {56--62} (\bibinfo {year} {2008})}\BibitemShut {NoStop}%
\bibitem [{\citenamefont {Marchetti}(2012)}]{marchetti3}%
  \BibitemOpen
  \bibfield  {author} {\bibinfo {author} {\bibfnamefont {M.~C.}\ \bibnamefont
  {Marchetti}},\ }\bibfield  {title} {\enquote {\bibinfo {title} {Spontaneous
  fows and self-propelled drops},}\ }\href@noop {} {\bibfield  {journal}
  {\bibinfo  {journal} {Nature}\ }\textbf {\bibinfo {volume} {491}},\ \bibinfo
  {pages} {340--341} (\bibinfo {year} {2012})}\BibitemShut {NoStop}%
\bibitem [{\citenamefont {Rajabi}\ \emph {et~al.}(2021)\citenamefont {Rajabi},
  \citenamefont {Baza}, \citenamefont {Turiv},\ and\ \citenamefont
  {Lavrentovich}}]{lavrentovich}%
  \BibitemOpen
  \bibfield  {author} {\bibinfo {author} {\bibfnamefont {M.}~\bibnamefont
  {Rajabi}}, \bibinfo {author} {\bibfnamefont {H.}~\bibnamefont {Baza}},
  \bibinfo {author} {\bibfnamefont {T.}~\bibnamefont {Turiv}}, \ and\ \bibinfo
  {author} {\bibfnamefont {O.~D.}\ \bibnamefont {Lavrentovich}},\ }\bibfield
  {title} {\enquote {\bibinfo {title} {Directional self-locomotion of active
  droplets enabled by nematic environment},}\ }\href@noop {} {\bibfield
  {journal} {\bibinfo  {journal} {Nature Physics}\ }\textbf {\bibinfo {volume}
  {17}},\ \bibinfo {pages} {260--266} (\bibinfo {year} {2021})}\BibitemShut
  {NoStop}%
\bibitem [{\citenamefont {Tiribocchi}\ \emph
  {et~al.}(2021{\natexlab{a}})\citenamefont {Tiribocchi}, \citenamefont
  {Montessori}, \citenamefont {Lauricella}, \citenamefont {Bonaccorso},
  \citenamefont {Succi}, \citenamefont {Aime}, \citenamefont {Milani},\ and\
  \citenamefont {Weitz}}]{tiribocchi2}%
  \BibitemOpen
  \bibfield  {author} {\bibinfo {author} {\bibfnamefont {A.}~\bibnamefont
  {Tiribocchi}}, \bibinfo {author} {\bibfnamefont {A.}~\bibnamefont
  {Montessori}}, \bibinfo {author} {\bibfnamefont {M.}~\bibnamefont
  {Lauricella}}, \bibinfo {author} {\bibfnamefont {F.}~\bibnamefont
  {Bonaccorso}}, \bibinfo {author} {\bibfnamefont {S.}~\bibnamefont {Succi}},
  \bibinfo {author} {\bibfnamefont {S.}~\bibnamefont {Aime}}, \bibinfo {author}
  {\bibfnamefont {M.}~\bibnamefont {Milani}}, \ and\ \bibinfo {author}
  {\bibfnamefont {D.}~\bibnamefont {Weitz}},\ }\bibfield  {title} {\enquote
  {\bibinfo {title} {The vortex-drive dyanamics of droplets within droplets},}\
  }\href {\doibase https://doi.org/10.1038/s41467-020-20364-0} {\bibfield
  {journal} {\bibinfo  {journal} {Nature Commun.}\ }\textbf {\bibinfo {volume}
  {12}},\ \bibinfo {pages} {82} (\bibinfo {year}
  {2021}{\natexlab{a}})}\BibitemShut {NoStop}%
\bibitem [{\citenamefont {Tiribocchi}\ \emph
  {et~al.}(2021{\natexlab{b}})\citenamefont {Tiribocchi}, \citenamefont
  {Montessori}, \citenamefont {Durve}, \citenamefont {Bonaccorso},
  \citenamefont {Lauricella},\ and\ \citenamefont {Succi}}]{tir_pre}%
  \BibitemOpen
  \bibfield  {author} {\bibinfo {author} {\bibfnamefont {A.}~\bibnamefont
  {Tiribocchi}}, \bibinfo {author} {\bibfnamefont {A.}~\bibnamefont
  {Montessori}}, \bibinfo {author} {\bibfnamefont {M.}~\bibnamefont {Durve}},
  \bibinfo {author} {\bibfnamefont {F.}~\bibnamefont {Bonaccorso}}, \bibinfo
  {author} {\bibfnamefont {M.}~\bibnamefont {Lauricella}}, \ and\ \bibinfo
  {author} {\bibfnamefont {S.}~\bibnamefont {Succi}},\ }\bibfield  {title}
  {\enquote {\bibinfo {title} {Dynamics of polydisperse multiple emulsions in
  microfluidic channels},}\ }\href {\doibase
  https://doi.org/10.1103/PhysRevE.104.065112} {\bibfield  {journal} {\bibinfo
  {journal} {Phys. Rev. E}\ }\textbf {\bibinfo {volume} {104}},\ \bibinfo
  {pages} {065112} (\bibinfo {year} {2021}{\natexlab{b}})}\BibitemShut
  {NoStop}%
\bibitem [{\citenamefont {Tiribocchi}\ \emph {et~al.}(2020)\citenamefont
  {Tiribocchi}, \citenamefont {Montessori}, \citenamefont {Aime}, \citenamefont
  {Milani}, \citenamefont {Lauricella}, \citenamefont {Succi},\ and\
  \citenamefont {Weitz}}]{pof1}%
  \BibitemOpen
  \bibfield  {author} {\bibinfo {author} {\bibfnamefont {A.}~\bibnamefont
  {Tiribocchi}}, \bibinfo {author} {\bibfnamefont {A.}~\bibnamefont
  {Montessori}}, \bibinfo {author} {\bibfnamefont {S.}~\bibnamefont {Aime}},
  \bibinfo {author} {\bibfnamefont {S.}~\bibnamefont {Milani}}, \bibinfo
  {author} {\bibfnamefont {M.}~\bibnamefont {Lauricella}}, \bibinfo {author}
  {\bibfnamefont {S.}~\bibnamefont {Succi}}, \ and\ \bibinfo {author}
  {\bibfnamefont {D.~A.}\ \bibnamefont {Weitz}},\ }\bibfield  {title} {\enquote
  {\bibinfo {title} {Novel nonequilibrium steady states in multiple
  emulsions},}\ }\href {\doibase https://doi.org/10.1063/1.5134901} {\bibfield
  {journal} {\bibinfo  {journal} {Physics of Fluids}\ }\textbf {\bibinfo
  {volume} {32}},\ \bibinfo {pages} {017102} (\bibinfo {year}
  {2020})}\BibitemShut {NoStop}%
\bibitem [{\citenamefont {Tiribocchi}\ \emph
  {et~al.}(2021{\natexlab{c}})\citenamefont {Tiribocchi}, \citenamefont
  {Montessori}, \citenamefont {Bonaccorso}, \citenamefont {Lauricella},\ and\
  \citenamefont {Succi}}]{pof2}%
  \BibitemOpen
  \bibfield  {author} {\bibinfo {author} {\bibfnamefont {A.}~\bibnamefont
  {Tiribocchi}}, \bibinfo {author} {\bibfnamefont {A.}~\bibnamefont
  {Montessori}}, \bibinfo {author} {\bibfnamefont {S.}~\bibnamefont
  {Bonaccorso}}, \bibinfo {author} {\bibfnamefont {M.}~\bibnamefont
  {Lauricella}}, \ and\ \bibinfo {author} {\bibfnamefont {S.}~\bibnamefont
  {Succi}},\ }\bibfield  {title} {\enquote {\bibinfo {title} {Shear dynamics of
  polydisperse double emulsions},}\ }\href {\doibase
  https://doi.org/10.1063/5.0046446} {\bibfield  {journal} {\bibinfo  {journal}
  {Physics of Fluids}\ }\textbf {\bibinfo {volume} {33}},\ \bibinfo {pages}
  {047105} (\bibinfo {year} {2021}{\natexlab{c}})}\BibitemShut {NoStop}%
\bibitem [{\citenamefont {de~Gennes}\ and\ \citenamefont
  {Prost}(1993)}]{degennes}%
  \BibitemOpen
  \bibfield  {author} {\bibinfo {author} {\bibfnamefont {P.~G.}\ \bibnamefont
  {de~Gennes}}\ and\ \bibinfo {author} {\bibfnamefont {J.}~\bibnamefont
  {Prost}},\ }\href@noop {} {\emph {\bibinfo {title} {The physics of liquid
  crystals}}}\ (\bibinfo  {publisher} {Oxford University Press, 2nd ed., ISBN:
  9780198517856},\ \bibinfo {year} {1993})\BibitemShut {NoStop}%
\bibitem [{\citenamefont {Succi}(2018)}]{succi}%
  \BibitemOpen
  \bibfield  {author} {\bibinfo {author} {\bibfnamefont {S.}~\bibnamefont
  {Succi}},\ }\href@noop {} {\emph {\bibinfo {title} {The Lattice {B}oltzmann
  equation: For complex states of flowing matter}}}\ (\bibinfo  {publisher}
  {Oxford University Press},\ \bibinfo {year} {2018})\BibitemShut {NoStop}%
\bibitem [{\citenamefont {Benzi}\ \emph {et~al.}(1992)\citenamefont {Benzi},
  \citenamefont {Succi},\ and\ \citenamefont {Vergassola}}]{vergassola}%
  \BibitemOpen
  \bibfield  {author} {\bibinfo {author} {\bibfnamefont {R.}~\bibnamefont
  {Benzi}}, \bibinfo {author} {\bibfnamefont {S.}~\bibnamefont {Succi}}, \ and\
  \bibinfo {author} {\bibfnamefont {M.}~\bibnamefont {Vergassola}},\ }\bibfield
   {title} {\enquote {\bibinfo {title} {The lattice boltzmann equation: theory
  and applications},}\ }\href {\doibase
  https://doi.org/10.1016/0370-1573(92)90090-M} {\bibfield  {journal} {\bibinfo
   {journal} {Physics Reports}\ }\textbf {\bibinfo {volume} {222}},\ \bibinfo
  {pages} {145--197} (\bibinfo {year} {1992})}\BibitemShut {NoStop}%
\bibitem [{\citenamefont {Higuera}\ \emph {et~al.}(1989)\citenamefont
  {Higuera}, \citenamefont {Succi},\ and\ \citenamefont {Benzi}}]{higuera}%
  \BibitemOpen
  \bibfield  {author} {\bibinfo {author} {\bibfnamefont {F.~J.}\ \bibnamefont
  {Higuera}}, \bibinfo {author} {\bibfnamefont {S.}~\bibnamefont {Succi}}, \
  and\ \bibinfo {author} {\bibfnamefont {R.}~\bibnamefont {Benzi}},\ }\bibfield
   {title} {\enquote {\bibinfo {title} {Lattice gas dynamics with enhanced
  collisions},}\ }\href {\doibase 10.1209/0295-5075/9/4/008} {\bibfield
  {journal} {\bibinfo  {journal} {Europhys. Lett.}\ }\textbf {\bibinfo {volume}
  {9}},\ \bibinfo {pages} {345} (\bibinfo {year} {1989})}\BibitemShut {NoStop}%
\bibitem [{\citenamefont {Marenduzzo}\ \emph {et~al.}(2007)\citenamefont
  {Marenduzzo}, \citenamefont {Orlandini}, \citenamefont {Cates},\ and\
  \citenamefont {Yeomans}}]{marenduzzo_pre}%
  \BibitemOpen
  \bibfield  {author} {\bibinfo {author} {\bibfnamefont {D.}~\bibnamefont
  {Marenduzzo}}, \bibinfo {author} {\bibfnamefont {E.}~\bibnamefont
  {Orlandini}}, \bibinfo {author} {\bibfnamefont {M.~E.}\ \bibnamefont
  {Cates}}, \ and\ \bibinfo {author} {\bibfnamefont {J.~M.}\ \bibnamefont
  {Yeomans}},\ }\bibfield  {title} {\enquote {\bibinfo {title} {Steady-state
  hydrodynamic instabilities of active liquid crystals: Hybrid lattice
  boltzmann simulations},}\ }\href {\doibase
  https://doi.org/10.1103/PhysRevE.76.031921} {\bibfield  {journal} {\bibinfo
  {journal} {Phys. Rev. E}\ }\textbf {\bibinfo {volume} {76}},\ \bibinfo
  {pages} {031921} (\bibinfo {year} {2007})}\BibitemShut {NoStop}%
\bibitem [{\citenamefont {Tjhung}\ \emph {et~al.}(2015)\citenamefont {Tjhung},
  \citenamefont {Tiribocchi}, \citenamefont {Marenduzzo},\ and\ \citenamefont
  {Cates}}]{tjhung2}%
  \BibitemOpen
  \bibfield  {author} {\bibinfo {author} {\bibfnamefont {E.}~\bibnamefont
  {Tjhung}}, \bibinfo {author} {\bibfnamefont {A.}~\bibnamefont {Tiribocchi}},
  \bibinfo {author} {\bibfnamefont {D.}~\bibnamefont {Marenduzzo}}, \ and\
  \bibinfo {author} {\bibfnamefont {M.~E.}\ \bibnamefont {Cates}},\ }\bibfield
  {title} {\enquote {\bibinfo {title} {A minimal physical model captures the
  shapes of crawling cells},}\ }\href {\doibase 10.1038/ncomms6420.} {\bibfield
   {journal} {\bibinfo  {journal} {Nature Commun.}\ }\textbf {\bibinfo {volume}
  {6}},\ \bibinfo {pages} {5420} (\bibinfo {year} {2015})}\BibitemShut
  {NoStop}%
\bibitem [{\citenamefont {Coupier}\ \emph {et~al.}(2012)\citenamefont
  {Coupier}, \citenamefont {Farutin}, \citenamefont {Minetti}, \citenamefont
  {Podgorski},\ and\ \citenamefont {Misbah}}]{coupier}%
  \BibitemOpen
  \bibfield  {author} {\bibinfo {author} {\bibfnamefont {G.}~\bibnamefont
  {Coupier}}, \bibinfo {author} {\bibfnamefont {A.}~\bibnamefont {Farutin}},
  \bibinfo {author} {\bibfnamefont {C.}~\bibnamefont {Minetti}}, \bibinfo
  {author} {\bibfnamefont {T.}~\bibnamefont {Podgorski}}, \ and\ \bibinfo
  {author} {\bibfnamefont {C.}~\bibnamefont {Misbah}},\ }\bibfield  {title}
  {\enquote {\bibinfo {title} {Shape diagram of vesicles in poiseuille flow},}\
  }\href@noop {} {\bibfield  {journal} {\bibinfo  {journal} {Phys. Rev. Lett.}\
  }\textbf {\bibinfo {volume} {108}},\ \bibinfo {pages} {178106} (\bibinfo
  {year} {2012})}\BibitemShut {NoStop}%
\bibitem [{\citenamefont {Kruse}\ \emph {et~al.}(2004)\citenamefont {Kruse},
  \citenamefont {Joanny}, \citenamefont {J\"ulicher}, \citenamefont {Prost},\
  and\ \citenamefont {Sekimoto}}]{kruse}%
  \BibitemOpen
  \bibfield  {author} {\bibinfo {author} {\bibfnamefont {K.}~\bibnamefont
  {Kruse}}, \bibinfo {author} {\bibfnamefont {J.-F.}\ \bibnamefont {Joanny}},
  \bibinfo {author} {\bibfnamefont {F.}~\bibnamefont {J\"ulicher}}, \bibinfo
  {author} {\bibfnamefont {J.}~\bibnamefont {Prost}}, \ and\ \bibinfo {author}
  {\bibfnamefont {K.}~\bibnamefont {Sekimoto}},\ }\bibfield  {title} {\enquote
  {\bibinfo {title} {Asters, vortices, and rotating spirals in active gels of
  polar filaments},}\ }\href {\doibase
  https://doi.org/10.1103/PhysRevLett.92.078101} {\bibfield  {journal}
  {\bibinfo  {journal} {Phys. Rev. Lett.}\ }\textbf {\bibinfo {volume} {92}},\
  \bibinfo {pages} {078101} (\bibinfo {year} {2004})}\BibitemShut {NoStop}%
\bibitem [{\citenamefont {J\"ulicher}\ \emph {et~al.}(2007)\citenamefont
  {J\"ulicher}, \citenamefont {Kruse}, \citenamefont {Prost},\ and\
  \citenamefont {Joanny}}]{julicher}%
  \BibitemOpen
  \bibfield  {author} {\bibinfo {author} {\bibfnamefont {F.}~\bibnamefont
  {J\"ulicher}}, \bibinfo {author} {\bibfnamefont {K.}~\bibnamefont {Kruse}},
  \bibinfo {author} {\bibfnamefont {J.}~\bibnamefont {Prost}}, \ and\ \bibinfo
  {author} {\bibfnamefont {J.-F.}\ \bibnamefont {Joanny}},\ }\bibfield  {title}
  {\enquote {\bibinfo {title} {Active behavior of the cytoskeleton},}\ }\href
  {\doibase https://doi.org/10.1016/j.physrep.2007.02.018} {\bibfield
  {journal} {\bibinfo  {journal} {Physics Reports}\ }\textbf {\bibinfo {volume}
  {449}},\ \bibinfo {pages} {3--28} (\bibinfo {year} {2007})}\BibitemShut
  {NoStop}%
\bibitem [{\citenamefont {Yang}\ \emph {et~al.}(2001)\citenamefont {Yang},
  \citenamefont {Fu},\ and\ \citenamefont {Lin}}]{yang}%
  \BibitemOpen
  \bibfield  {author} {\bibinfo {author} {\bibfnamefont {R.~J.}\ \bibnamefont
  {Yang}}, \bibinfo {author} {\bibfnamefont {L.~M.}\ \bibnamefont {Fu}}, \ and\
  \bibinfo {author} {\bibfnamefont {Y.~C.}\ \bibnamefont {Lin}},\ }\bibfield
  {title} {\enquote {\bibinfo {title} {Electroosmotic flow in microchannels},}\
  }\href@noop {} {\bibfield  {journal} {\bibinfo  {journal} {Jour. Coll. and
  Int. Sci.}\ }\textbf {\bibinfo {volume} {239}},\ \bibinfo {pages} {98--105}
  (\bibinfo {year} {2001})}\BibitemShut {NoStop}%
\bibitem [{\citenamefont {Tang}\ \emph {et~al.}(2009)\citenamefont {Tang},
  \citenamefont {Li}, \citenamefont {He},\ and\ \citenamefont {Tao}}]{tao}%
  \BibitemOpen
  \bibfield  {author} {\bibinfo {author} {\bibfnamefont {G.~H.}\ \bibnamefont
  {Tang}}, \bibinfo {author} {\bibfnamefont {X.~F.}\ \bibnamefont {Li}},
  \bibinfo {author} {\bibfnamefont {Y.~L.}\ \bibnamefont {He}}, \ and\ \bibinfo
  {author} {\bibfnamefont {W.~Q.}\ \bibnamefont {Tao}},\ }\bibfield  {title}
  {\enquote {\bibinfo {title} {Electroosmotic flow of non-newtonian fluid in
  microchannels},}\ }\href@noop {} {\bibfield  {journal} {\bibinfo  {journal}
  {J. Non-Newt. Fluid Mech.}\ }\textbf {\bibinfo {volume} {157}},\ \bibinfo
  {pages} {133--137} (\bibinfo {year} {2009})}\BibitemShut {NoStop}%
\bibitem [{\citenamefont {Shit}\ \emph {et~al.}(2016)\citenamefont {Shit},
  \citenamefont {Mondal}, \citenamefont {Sinha},\ and\ \citenamefont
  {Kundu}}]{kundu}%
  \BibitemOpen
  \bibfield  {author} {\bibinfo {author} {\bibfnamefont {G.~C.}\ \bibnamefont
  {Shit}}, \bibinfo {author} {\bibfnamefont {A.}~\bibnamefont {Mondal}},
  \bibinfo {author} {\bibfnamefont {A.}~\bibnamefont {Sinha}}, \ and\ \bibinfo
  {author} {\bibfnamefont {P.~K.}\ \bibnamefont {Kundu}},\ }\bibfield  {title}
  {\enquote {\bibinfo {title} {Two-layer electro-osmotic flow and heat transfer
  in a hydrophobic micro-channel with fluid-solid interfacial slip and zeta
  potential difference},}\ }\href {\doibase https://doi.org/10.1063/1.1744102}
  {\bibfield  {journal} {\bibinfo  {journal} {Coll. and Surf. A: Physicoch. and
  Engi. Asp.}\ }\textbf {\bibinfo {volume} {506}},\ \bibinfo {pages} {535--549}
  (\bibinfo {year} {2016})}\BibitemShut {NoStop}%
\bibitem [{\citenamefont {Ramos}\ \emph {et~al.}(2020)\citenamefont {Ramos},
  \citenamefont {Cordero},\ and\ \citenamefont {Soto}}]{soto_softmatter}%
  \BibitemOpen
  \bibfield  {author} {\bibinfo {author} {\bibfnamefont {G.}~\bibnamefont
  {Ramos}}, \bibinfo {author} {\bibfnamefont {M.~L.}\ \bibnamefont {Cordero}},
  \ and\ \bibinfo {author} {\bibfnamefont {R.}~\bibnamefont {Soto}},\
  }\bibfield  {title} {\enquote {\bibinfo {title} {Bacteria driving
  droplets},}\ }\href@noop {} {\bibfield  {journal} {\bibinfo  {journal} {Soft
  Matter}\ }\textbf {\bibinfo {volume} {16}},\ \bibinfo {pages} {1359--1365}
  (\bibinfo {year} {2020})}\BibitemShut {NoStop}%
\bibitem [{\citenamefont {Mi\~{n}o}\ \emph {et~al.}(2011)\citenamefont
  {Mi\~{n}o}, \citenamefont {Mallouk}, \citenamefont {Darnige}, \citenamefont
  {Hoyos}, \citenamefont {Dauchet}, \citenamefont {Dunstan}, \citenamefont
  {Soto}, \citenamefont {Wang}, \citenamefont {Rousselet},\ and\ \citenamefont
  {Clement}}]{soto_prl}%
  \BibitemOpen
  \bibfield  {author} {\bibinfo {author} {\bibfnamefont {G.}~\bibnamefont
  {Mi\~{n}o}}, \bibinfo {author} {\bibfnamefont {T.~E.}\ \bibnamefont
  {Mallouk}}, \bibinfo {author} {\bibfnamefont {T.}~\bibnamefont {Darnige}},
  \bibinfo {author} {\bibfnamefont {M.}~\bibnamefont {Hoyos}}, \bibinfo
  {author} {\bibfnamefont {J.}~\bibnamefont {Dauchet}}, \bibinfo {author}
  {\bibfnamefont {J.}~\bibnamefont {Dunstan}}, \bibinfo {author} {\bibfnamefont
  {R.}~\bibnamefont {Soto}}, \bibinfo {author} {\bibfnamefont {Y.}~\bibnamefont
  {Wang}}, \bibinfo {author} {\bibfnamefont {A.}~\bibnamefont {Rousselet}}, \
  and\ \bibinfo {author} {\bibfnamefont {E.}~\bibnamefont {Clement}},\
  }\bibfield  {title} {\enquote {\bibinfo {title} {Enhanced diffusion due to
  active swimmers at a solid surface},}\ }\href@noop {} {\bibfield  {journal}
  {\bibinfo  {journal} {Phys. Rev. Lett.}\ }\textbf {\bibinfo {volume} {106}},\
  \bibinfo {pages} {048102} (\bibinfo {year} {2011})}\BibitemShut {NoStop}%
\bibitem [{\citenamefont {Miller}\ \emph {et~al.}(2013)\citenamefont {Miller},
  \citenamefont {Carlton}, \citenamefont {Mushenheim},\ and\ \citenamefont
  {Abbott}}]{abbott}%
  \BibitemOpen
  \bibfield  {author} {\bibinfo {author} {\bibfnamefont {D.~S.}\ \bibnamefont
  {Miller}}, \bibinfo {author} {\bibfnamefont {R.~J.}\ \bibnamefont {Carlton}},
  \bibinfo {author} {\bibfnamefont {P.~C.}\ \bibnamefont {Mushenheim}}, \ and\
  \bibinfo {author} {\bibfnamefont {N.~L.}\ \bibnamefont {Abbott}},\ }\bibfield
   {title} {\enquote {\bibinfo {title} {Introduction to optical methods for
  characterizing liquid crystals at interfaces},}\ }\href@noop {} {\bibfield
  {journal} {\bibinfo  {journal} {Langmuir}\ }\textbf {\bibinfo {volume}
  {29}},\ \bibinfo {pages} {3154--3169} (\bibinfo {year} {2013})}\BibitemShut
  {NoStop}%
\bibitem [{\citenamefont {Schwarz-Linek}\ \emph {et~al.}(2012)\citenamefont
  {Schwarz-Linek}, \citenamefont {Valeriani}, \citenamefont {Cacciuto},
  \citenamefont {Cates}, \citenamefont {Marenduzzo}, \citenamefont {Morozov},\
  and\ \citenamefont {Poon}}]{linek}%
  \BibitemOpen
  \bibfield  {author} {\bibinfo {author} {\bibfnamefont {J.}~\bibnamefont
  {Schwarz-Linek}}, \bibinfo {author} {\bibfnamefont {C.}~\bibnamefont
  {Valeriani}}, \bibinfo {author} {\bibfnamefont {A.}~\bibnamefont {Cacciuto}},
  \bibinfo {author} {\bibfnamefont {M.~E.}\ \bibnamefont {Cates}}, \bibinfo
  {author} {\bibfnamefont {D.}~\bibnamefont {Marenduzzo}}, \bibinfo {author}
  {\bibfnamefont {A.~N.}\ \bibnamefont {Morozov}}, \ and\ \bibinfo {author}
  {\bibfnamefont {W.~C.~K.}\ \bibnamefont {Poon}},\ }\bibfield  {title}
  {\enquote {\bibinfo {title} {Phase separation and rotor self-assembly in
  active particle suspensions},}\ }\href {\doibase
  https://doi.org/10.1073/pnas.111633410} {\bibfield  {journal} {\bibinfo
  {journal} {Proc. Natl. Acad. Sci. USA}\ }\textbf {\bibinfo {volume} {109}},\
  \bibinfo {pages} {4052--4057} (\bibinfo {year} {2012})}\BibitemShut {NoStop}%
\bibitem [{\citenamefont {Guanglai}\ \emph {et~al.}(2011)\citenamefont
  {Guanglai}, \citenamefont {Bensson}, \citenamefont {Nisomova}, \citenamefont
  {Munger}, \citenamefont {Mahatumr}, \citenamefont {Tang}, \citenamefont
  {Maxey},\ and\ \citenamefont {Brun}}]{guanglai_pre}%
  \BibitemOpen
  \bibfield  {author} {\bibinfo {author} {\bibfnamefont {L.}~\bibnamefont
  {Guanglai}}, \bibinfo {author} {\bibfnamefont {J.}~\bibnamefont {Bensson}},
  \bibinfo {author} {\bibfnamefont {L.}~\bibnamefont {Nisomova}}, \bibinfo
  {author} {\bibfnamefont {D.}~\bibnamefont {Munger}}, \bibinfo {author}
  {\bibfnamefont {P.}~\bibnamefont {Mahatumr}}, \bibinfo {author}
  {\bibfnamefont {J.~X.}\ \bibnamefont {Tang}}, \bibinfo {author}
  {\bibfnamefont {M.~R.}\ \bibnamefont {Maxey}}, \ and\ \bibinfo {author}
  {\bibfnamefont {Y.~V.}\ \bibnamefont {Brun}},\ }\bibfield  {title} {\enquote
  {\bibinfo {title} {Accumulation of swimming bacteria near a solid surface},}\
  }\href {\doibase https://doi.org/10.1103/PhysRevE.84.041932} {\bibfield
  {journal} {\bibinfo  {journal} {Phys. Rev. E}\ }\textbf {\bibinfo {volume}
  {84}},\ \bibinfo {pages} {041932} (\bibinfo {year} {2011})}\BibitemShut
  {NoStop}%
\bibitem [{\citenamefont {Bonelli}\ \emph {et~al.}(2019)\citenamefont
  {Bonelli}, \citenamefont {Carenza}, \citenamefont {Gonnella}, \citenamefont
  {Marenduzzo}, \citenamefont {Orlanini},\ and\ \citenamefont
  {Tiribocchi}}]{sci_rep_bon}%
  \BibitemOpen
  \bibfield  {author} {\bibinfo {author} {\bibfnamefont {F.}~\bibnamefont
  {Bonelli}}, \bibinfo {author} {\bibfnamefont {L.~N.}\ \bibnamefont
  {Carenza}}, \bibinfo {author} {\bibfnamefont {G.}~\bibnamefont {Gonnella}},
  \bibinfo {author} {\bibfnamefont {D.}~\bibnamefont {Marenduzzo}}, \bibinfo
  {author} {\bibfnamefont {E.}~\bibnamefont {Orlanini}}, \ and\ \bibinfo
  {author} {\bibfnamefont {A.}~\bibnamefont {Tiribocchi}},\ }\bibfield  {title}
  {\enquote {\bibinfo {title} {Lamellar ordering, droplet formation and phase
  inversion in exotic active emulsions},}\ }\href {\doibase
  https://doi.org/10.1038/s41598-019-39190-6} {\bibfield  {journal} {\bibinfo
  {journal} {Sci. Rep.}\ }\textbf {\bibinfo {volume} {9}},\ \bibinfo {pages}
  {2801} (\bibinfo {year} {2019})}\BibitemShut {NoStop}%
\bibitem [{\citenamefont {Masaeli}\ \emph {et~al.}(2012)\citenamefont
  {Masaeli}, \citenamefont {Sollier}, \citenamefont {Amini}, \citenamefont
  {Mao}, \citenamefont {Camacho}, \citenamefont {Doshi}, \citenamefont
  {Mitragotri}, \citenamefont {Alexeev},\ and\ \citenamefont
  {Di~Carlo}}]{masaeli}%
  \BibitemOpen
  \bibfield  {author} {\bibinfo {author} {\bibfnamefont {M.}~\bibnamefont
  {Masaeli}}, \bibinfo {author} {\bibfnamefont {E.}~\bibnamefont {Sollier}},
  \bibinfo {author} {\bibfnamefont {H.}~\bibnamefont {Amini}}, \bibinfo
  {author} {\bibfnamefont {W.}~\bibnamefont {Mao}}, \bibinfo {author}
  {\bibfnamefont {K.}~\bibnamefont {Camacho}}, \bibinfo {author} {\bibfnamefont
  {N.}~\bibnamefont {Doshi}}, \bibinfo {author} {\bibfnamefont
  {S.}~\bibnamefont {Mitragotri}}, \bibinfo {author} {\bibfnamefont
  {A.}~\bibnamefont {Alexeev}}, \ and\ \bibinfo {author} {\bibfnamefont
  {D.}~\bibnamefont {Di~Carlo}},\ }\bibfield  {title} {\enquote {\bibinfo
  {title} {Continuous inertial focusing and separation of particles by
  shape},}\ }\href {\doibase 10.1103/PhysRevX.2.031017} {\bibfield  {journal}
  {\bibinfo  {journal} {Phys. Rev. X}\ }\textbf {\bibinfo {volume} {2}},\
  \bibinfo {pages} {031017} (\bibinfo {year} {2012})}\BibitemShut {NoStop}%
\bibitem [{\citenamefont {Stone}(1994)}]{stone3}%
  \BibitemOpen
  \bibfield  {author} {\bibinfo {author} {\bibfnamefont {H.~A.}\ \bibnamefont
  {Stone}},\ }\bibfield  {title} {\enquote {\bibinfo {title} {Dynamics of drop
  deformation and breakup in viscous fluids},}\ }\href@noop {} {\bibfield
  {journal} {\bibinfo  {journal} {Annu. Rev. Fluid Mech.}\ }\textbf {\bibinfo
  {volume} {26}},\ \bibinfo {pages} {65} (\bibinfo {year} {1994})}\BibitemShut
  {NoStop}%
\bibitem [{\citenamefont {Bentley}\ and\ \citenamefont {Leal}(1986)}]{bentley}%
  \BibitemOpen
  \bibfield  {author} {\bibinfo {author} {\bibfnamefont {B.~J.}\ \bibnamefont
  {Bentley}}\ and\ \bibinfo {author} {\bibfnamefont {L.~G.}\ \bibnamefont
  {Leal}},\ }\bibfield  {title} {\enquote {\bibinfo {title} {An experimental
  investigation of drop deformation and breakup in steady, two-dimensional
  linear flows},}\ }\href@noop {} {\bibfield  {journal} {\bibinfo  {journal}
  {J. Fluid. Mech.}\ }\textbf {\bibinfo {volume} {167}},\ \bibinfo {pages}
  {241} (\bibinfo {year} {1986})}\BibitemShut {NoStop}%
\end{thebibliography}%

\end{document}